\documentclass[pdflatex,sn-mathphys-num]{sn-jnl}

\usepackage{graphicx}%
\usepackage{multirow}%
\usepackage{amsmath,amssymb,amsfonts}%
\usepackage{amsthm}%
\usepackage{mathrsfs}%
\usepackage[title]{appendix}%
\usepackage{xcolor}%
\usepackage{textcomp}%
\usepackage{manyfoot}%
\usepackage{booktabs}%
\usepackage{algorithm}%
\usepackage{algorithmicx}%
\usepackage{algpseudocode}%
\usepackage{listings}%
\usepackage{nicefrac}
\usepackage{xurl}
\newcommand{\Sabs}[1]{\left\lceil #1 \right\rfloor}

\newcommand{\norm}[1]{\left|\left| #1 \right|\right|}

\newcommand{\sign}{\textup{sign}}

\theoremstyle{thmstyleone}%
\newtheorem{theorem}{Theorem}
\theoremstyle{thmstyletwo}%
\newtheorem{remark}{Remark}%
\newtheorem{assumption}{Assumption}

\theoremstyle{thmstylethree}%

\begin{document}

\title[Control of Human-Induced Seismicity in Underground Reservoirs Governed by a Nonlinear PDE-ODE System]{Control of Human-Induced Seismicity in Underground Reservoirs Governed by a Nonlinear PDE-ODE System}


\author[1]{\fnm{Diego} \sur{Guti\'errez-Oribio}}\email{diego.gutierrez@ensta.fr}

\author*[1]{\fnm{Ioannis} \sur{Stefanou}}\email{ioannis.stefanou@ensta.fr}

\affil[1]{\orgdiv{IMSIA (UMR 9219)}, \orgname{CNRS, EDF, ENSTA Paris, Institut Polytechnique de Paris}, \orgaddress{\street{828, Boulevard des Mar\'echaux}, \postcode{91762}, \state{Palaiseau}, \country{France}}}


\abstract{Induced seismicity caused by fluid extraction or injection in underground reservoirs is a major challenge for safe energy production and storage. This paper presents a robust output-feedback controller for induced seismicity mitigation in geological reservoirs described by a coupled PDE-ODE model. The used MIMO Super-Twisting controller is nonlinear and robust, producing a continuous control signal and requiring minimal model information, while accommodating parameter uncertainties and spatial heterogeneity. Two operational outputs are regulated simultaneously: regional pressures and seismicity rates computed over reservoir sub-regions. Closed-loop properties are established via bounds on the PDE-ODE solutions, the uncertain control coefficient, and the time derivative of the regional perturbations, yielding finite-time stability or input-to-state stability of the tracking errors, depending on the controller parameters. Inspired by the Groningen gas reservoir, the method is illustrated in two academic scenarios: gas production with a prescribed SR target and a conceptual comparison using gas production and CO$_2$ injection. Simulations demonstrate accurate tracking of pressure and seismicity targets across regions under significant parameter uncertainty, supporting safer reservoir operation while preserving production objectives.}




\maketitle

\section{Introduction}
\label{sec:introduction}

In the realm of complex dynamical systems, the interplay between time-dependent and spatially-dependent behaviours is challenging for both modelling and control. Ordinary Differential Equations (ODEs) effectively capture the dynamics of systems governed by a single independent variable, typically time, and are widely used in many fields such as physics, engineering, and economics \citep{b:Khalil2002}. Conversely, Partial Differential Equations (PDEs) allow for the description of systems where multiple independent variables, including spatial dimensions, significantly influence the behaviour of the system \citep{b:EvansPDE}. As a result, PDE-ODE systems usually emerge in the modelling of complex phenomena and processes that exhibit both temporal evolution and spatial interaction.

These kinds of systems are prevalent in various applications, ranging from continuum mechanics, heat transfer, electromagnetism \citep{b:zhong2016coupled,b:koch2022sliding,b:quarteroni2003analysis}, to seismicity control of underground reservoirs, which is studied here. The intricate structure of such systems necessitates non-trivial and sophisticated control strategies capable of addressing the dynamic coupling between the PDE and ODE components in a robust way. 

The motivation for the theoretical developments presented in this work
stems from the challenging problem of mitigating induced seismicity
caused by fluid injection and extraction in underground geological
reservoirs. Deep geothermal energy, carbon capture and storage, and
hydrogen storage have demonstrated significant potential to meet the
growing demands of the energy sector while reducing CO$_2$ emissions.
However, these techniques can inadvertently induce seismic events
\citep{b:10.1785/0220170112,b:Rubinstein-Mahani-2015,
b:10.1002/2016RG000542}. This concern has led to the closure of
several plants worldwide, \textit{e.g.},
\citep{b:Stey_LeMonde,b:Zastrow-2019,b:10.1785/gssrl.80.5.784}.
Although different mitigation strategies have been proposed, their
effectiveness remains limited, particularly when seismicity reduction
must be achieved while maintaining energy production and storage
\citep{b:Verdon-Bommer-2021,b:10.1093/gji/ggz058,
b:doi.org/10.1029/2020GL090648,
https://doi.org/10.1007/s00603-018-1467-4,
b:10.1785/0220180337,b:10.1093/gji/ggac416}.

In recent years, control theory has been applied to seismic
instabilities in specific, well-characterized, mature faults
\citep{b:Stefanou2019,b:https://doi.org/10.1029/2021JB023410,
b:Gutierrez-Tzortzopoulos-Stefanou-Plestan-2022,
b:Gutierrez-Orlov-Stefanou-Plestan-2023,
b:Gutierrez-Stefanou-Plestan-2024,
b:Gutierrez-Orlov-Plestan-Stefanou-VSS2022}.
These studies employed different control algorithms to slow the
release of accumulated energy at a rate far lower than that of
natural, uncontrolled seismic events. At the reservoir scale,
stress-based forecasting and operational optimization have been
studied for Groningen
\citep{b:SMITH2022117697,b:doi.org/10.1029/2023GL105455,
b:10.1785/0220230179,b:https://doi.org/10.1029/2024GL110139}.
Feedback control offers a different approach, where seismicity rate (SR) is used to adjust well rates
\citep{b:Gutierrez-Stefanou-2024}. However, due to mathematical
complexities, this approach considered simplified region-wise
SR models rather than point-wise dynamics, and regulated a
transformed SR output instead of the SR directly. These
limitations motivated the present work.

Other control techniques, such as optimal control, backstepping,
and observer-based feedback, have been developed for PDE and
PDE--ODE systems
\citep{b:fursikov1999optimal,b:martinez2018optimal,
b:deutscher2020output,b:tang2011state,b:zhao2021backstepping,
b:susto2010control,b:7307993}.
More recent results include adaptive observers \citep{b:Ji2023},
event-triggered control \citep{WangKrstic2023}, adaptive tracking
\citep{Xiao2025}, neural-operator-based feedback
\citep{JiangWang2025,JiangWang2026}, and sliding-mode control
\citep{b:koch2022sliding,b:wang2015sliding,OrlovEtAl2026,GuoEtAl2026}. However, optimal control
generally requires solving the PDE system, while backstepping
and observer-based approaches often require detailed model
information or auxiliary dynamics. These requirements increase
the computational complexity and may introduce spillover issues.

In this work, we consider a nonlinear PDE--ODE system in $n_d\in\{1,2,3\}$
spatial dimensions describing the evolution of the SR within a geological
reservoir. The SR is governed by a logistic-like nonlinear ODE,
which is one-way coupled, point-wise, with a diffusion PDE.
The latter is driven by localized fluid sources representing
injection and extraction processes. For this system, we develop
a robust output feedback controller that directly tracks
prescribed regional pressures and SR, while also allowing
constraints on the well fluxes. Contrary to the approaches
discussed above, the proposed control requires only measured
outputs and nominal control information, without an observer
for the distributed state. Although motivated by induced
seismicity, this control strategy may be applied to other PDE--ODE couplings, like multi-physics and multi-scale systems
\citep{b:deBuhanFrey2011,b:Rizzi2021,b:KURDYUMOV2015981,
b:helander2005collisional,b:GALLAY2022103387,
b:10.7554/eLife.37836}.

The proposed controller is based on a Multi-Input Multi-Output
(MIMO) Super-Twisting Algorithm. It generates continuous control signals and achieves finite-time
or exponential tracking, depending on the control parameters,
under the stated uncertainty conditions. The main mathematical
challenge is to establish the boundedness properties needed
to apply the MIMO Super-Twisting stability results to the
infinite-dimensional PDE--ODE system. For this purpose, we
first establish bounds on the diffusion equation and its
time derivative, which are then extended to the nonlinear
SR dynamics. These results allow us to bound the uncertain
control coefficient and the time derivative of the regional
perturbations, leading to the desired tracking results.

To illustrate the proposed approach, we consider the Groningen
gas reservoir, which was closed in 2024 after approximately
60 years of operation due to unacceptable levels of induced
seismicity \citep{b:Groningen1,b:Groningen2}.
Numerical simulations using a reduced model inspired by established works in
Groningen \citep{b:https://doi.org/10.1029/2020JB020013,doi:10.1144/SP528-2022-169,b:doi.org/10.1029/2023GL105455,BourneOates2017} illustrate the control performance in two academic
scenarios. In the first, the controller tracks prescribed
regional pressures and the background SR while satisfying
weighted production-demand constraints inspired by historical
extraction. In the second, gas production and a conceptual CO$_2$
injection are considered to illustrate the additional
control possibilities. 

The structure of this paper is outlined as follows. Section \ref{sec:problem} introduces the underlying PDE-ODE mathematical model and defines the control objectives. In Section \ref{sec:control}, the proposed robust output feedback controller is designed and mathematically proved. Section \ref{sec:simulations} illustrates the control with simulations of a reduced model fitted to Groningen observations. We provide in the same Section more details about the physics of the controlled PDE-ODE system and we discuss its limitations. Finally, Section \ref{sec:conclusion} provides concluding remarks and summarizes the key findings of the study.

\noindent \textbf{Notation:} We denote by $\|\cdot\|$ the Euclidean norm on $\mathbb{R}^n$. Let $\mathbb{I}_n$ be the $n\times n$ identity matrix and $0_{n\times m}$ the $n\times m$ zero matrix. The spectral norm of a matrix $A\in\mathbb{R}^{n\times m}$ is $\|A\|:=\sqrt{\lambda_{\max}(A^{\top}A)}$, where $\lambda_{\max}(\cdot)$ denotes the largest eigenvalue. A column vector $\Phi=[\phi_1,\dots,\phi_m]^{\top}$ is also denoted by $\Phi=[\phi_i]_{i=1}^m$, and a diagonal matrix $A=\mathrm{diag}(a_1,\dots,a_m)$ by $A=\mathrm{diag}(a_i)_{i=1}^m$. The partial time derivative of $u(x,t)$ is denoted by $u_t=\partial u/\partial t$, and the total time derivative by $\dot u=du/dt$. The gradient is $\nabla u=[\partial u/\partial x_1,\dots,\partial u/\partial x_{n_d}]^{\top}$, and the Laplacian is $\nabla^2 u=\nabla\cdot(\nabla u)=\sum_{i=1}^{n_d} \nicefrac{\partial^2 u}{\partial x_i^2}$. We denote by $L^p(U)$ the space of functions whose
$p$-th power is integrable over $U$, with $L^\infty(U)$
denoting the space of essentially bounded functions.
The Sobolev space $W^{k,p}(U)$ consists of functions
in $L^p(U)$ whose weak derivatives up to order $k$
also belong to $L^p(U)$. We write $H^k(U)=W^{k,2}(U)$. For functions in $L^2(U)$, we use the inner product
$\langle f,g\rangle=\int_U fg\,dV$ and the norm
$\|f\|_2=\|f\|_{L^2(U)}$.

\section{Problem Statement}
\label{sec:problem}

Let us introduce the cascade model describing induced seismicity in underground reservoirs driven by fluid injection and extraction. We begin with the diffusion system, written as
\begin{equation}
\begin{split}
  u_{t}(x,t) &= -\frac{1}{\beta} \nabla \cdot q(x,t)+ \frac{1}{\beta} \sum_{i=1}^{n} \mathcal{B}_i(x) Q_i(t),\\
  q(x,t) &= -\frac{k(x)}{\eta(x)} \nabla u(x,t), \\
  q(x,t) \cdot \hat{e} &= 0 \quad \forall \quad x \in \partial V,
\end{split}
\label{eq:diff}
\end{equation}
where $u(x,t)$ is the solution representing the change in pore pressure of the reservoir fluid, and $u(x,0) = u^0(x)$ its initial condition. $q(x,t)$ is a flux term according to Darcy's law (second equation in \eqref{eq:diff}). $x \in V$ represents the spatial variable and $t \in T$, $T:=[0, \infty)$ the time variable. $V\subset\mathbb{R}^{n_d}$, with $n_d\in\{1,2,3\}$, is a bounded, connected Lipschitz domain with Lipschitz boundary $\partial V$. Neumann boundary conditions are considered (undrained) where $\hat{e}$ is a unit normal vector to $\partial V$. $\beta>0$ is a system parameter, and $k(x) \in \mathbb{R}^{n_d\times n_d}$ and $\eta(x)$ are system functions that depend on the space variable. They represent the combined compressibility of the fluid and porous rock, the permeability matrix of the host rock, and the dynamic viscosity of the fluid, respectively. $Q(t) \in \mathbb{R}^{n}$, $Q(t)=[Q_{1}(t),...,Q_{n}(t)]^T$, are flux sources (inputs) applied through the coefficients $\mathcal{B}_i(x)$ defined as
\begin{equation}
\begin{split}
  \mathcal{B}_i(x) &= \left\{ \begin{array}{c}
  \frac{1}{|V_i^*|} \quad \textup{if} \quad x \in V_i^* \\ 
  \hspace{4pt} 0 \hspace{14pt} \textup{if} \quad x \notin V_i^*
  \end{array} \right. , \quad 
   i = 1,...,n, \quad V_i^* \subset V.
\end{split}
\label{eq:B}
\end{equation}
Here $V_i^*$ denotes the well source region associated with well $i$, that is, the part of the reservoir where the flux $Q_i(t)$ is distributed. Its $n_d$-dimensional measure is denoted by $|V_i^*|$. Note that $\mathcal{B}_i(x)\in L^2(V)$ ($\|\mathcal{B}_i\|_{L^2(V)}=1/\sqrt{|V_i^*|}$) and $\int_V\mathcal{B}_i(x)\,dV=\int_{V_i^*}\mathcal{B}_i(x)\,dV=1$ for all $i=1,\ldots,n$. Furthermore, $\mathcal{B}_i(x)$ tends to a Dirac delta distribution when $V_i^*$ shrinks to a point. The flux $Q_i(t)$ is positive for injection and negative for extraction.

Let the next ODE system be applied in a cascade connection with the diffusion equation as
\begin{equation}
\begin{split}
  R_t(x,t) = R(x,t) \Big\{s\gamma_1(x) u_t(x,t) - \gamma_2(x,t) \left[R(x,t)-R^*(x) \right] \Big\},
\end{split}
\label{eq:SR}
\end{equation}
where $R(x,t)$ expresses the seismicity-rate (SR) density at point $x$ and time $t$, with initial condition $R(x,0) = R^0(x) > 0$. The SR can only take positive values, \textit{i.e.}, $R(x,t)>0$ $\forall$ $(x,t) \in (V \times T)$. $\gamma_1(x)$ and $\gamma_2(x,t)$ are system functions that can depend on the space and time variables. $R^*(x) \in L^{2}(V)$ is the positive background SR density in the absence of operations, with the same units as $R(x,t)$. $u_t(x,t)$ is the input of this system, which is the time derivative of the solution of the diffusion system \eqref{eq:diff}. The sign $s\in\{-1,1\}$ is known and constant throughout the reservoir and in time. 

System \eqref{eq:diff}--\eqref{eq:SR} can be used to model the SR in underground reservoirs, \textit{i.e.}, the number of seismic events due to fluid injection and/or extraction. Section \ref{sec:simulations} provides more details about the physical interpretation of the system \eqref{eq:diff}--\eqref{eq:SR}. Similar PDE--ODE couplings also arise in a variety of challenging applications, including structural mechanics, mechanical metamaterials, combustion and plasma physics, tumor growth, and neuroscience, among many others \citep{b:deBuhanFrey2011,b:Rizzi2021,b:KURDYUMOV2015981,b:helander2005collisional,b:GALLAY2022103387,b:10.7554/eLife.37836}.

The objective of this work is to design the control input $Q(t)$ of system \eqref{eq:diff} so that region-averaged pressure and seismicity-rate outputs track prescribed references, \textit{i.e.}, to drive the outputs $y_u \in \mathbb{R}^{m_u}$, $y_u=[y_{u_1},...,y_{u_{m_u}}]^T$, and $y_R \in \mathbb{R}^{m_R}$, $y_R=[y_{R_1},...,y_{R_{m_R}}]^T$, with components
\begin{equation}
\begin{split}
  y_{u_i}(t) = \frac{1}{V_{u_i}} \int_{V_{u_i}} u(x,t) \, dV, \quad &V_{u_i} \subset V, \quad i = 1,...,m_u, \\
  y_{R_i}(t) = \frac{1}{V_{R_i}} \int_{V_{R_i}} R(x,t) \, dV, \quad &V_{R_i} \subset V, \quad i = 1,...,m_R,
\end{split}
  \label{eq:output}
\end{equation}
to desired references $r_u(t) \in \mathbb{R}^{m_u}$, $r_u(t)=[r_{u_1}(t),...,r_{u_{m_u}}(t)]^T$, and $r_R(t) \in \mathbb{R}^{m_R}$, $r_R(t)=[r_{R_1}(t),...,r_{R_{m_R}}(t)]^T$, $r_{R_i}>0$, respectively. 

The control design will be performed under the following assumptions for system \eqref{eq:diff}--\eqref{eq:SR}:

\begin{assumption}\label{A1}:
The actuation of the plant \eqref{eq:diff} implements the control law \eqref{eq:Q} or \eqref{eq:Qr}, with bounded flux sources. Consequently, there exists a constant \(0<L_Q< \infty\) such that
\begin{equation}
  \|Q(t)\| \leq L_Q, \quad \forall\, t \in T.
  \label{eq:Q_bounding}
\end{equation}
Moreover, the accumulated total well flux satisfies
\begin{equation}
 \sup_{t\geq0}\left|\int_0^t\sum_{j=1}^n Q_j(s)\,ds\right|\leq L_{\mathrm{net}}<\infty.
 \label{eq:initial_net_flux}
\end{equation}
\end{assumption}
\begin{assumption}\label{A2}:
The references to be followed, $r_u(t)$ and $r_R(t)$, are designed to fulfil
\begin{equation}
\begin{split}
  \norm{{r}_u(t)} \leq L_{{r}_u}, \quad \norm{\dot{r}_u(t)} &\leq L_{\dot{r}_u}, \quad \norm{\ddot{r}_u(t)} \leq L_{\ddot{r}_u}, \\
  \norm{{r}_R(t)} \leq L_{{r}_R}, \quad \norm{\dot{r}_R(t)} &\leq L_{\dot{r}_R}, \quad \norm{\ddot{r}_R(t)} \leq L_{\ddot{r}_R},
\end{split}
  \label{eq:ref}
\end{equation}
for all $t \in T$.
\end{assumption}
\begin{assumption}\label{A3}:
The system functions are uncertain, with $k(x)=k(x)^T$ and
\begin{equation}
\begin{split}
  0<k^m,\quad k^m|\xi|^2\leq \xi^T k(x)\xi\leq k^M|\xi|^2, \quad
  &0<\eta^m\leq\eta(x)\leq\eta^M,\\
  0<\gamma_1^m\leq\gamma_1(x)\leq\gamma_1^M, \quad
  &0<\gamma_2^m\leq\gamma_2(x,t)\leq\gamma_2^M,\\
  0<R_m^*\leq R^*(x)\leq R_M^*,\quad
  &|\gamma_{2,t}(x,t)|\leq L_{\gamma_{2,t}},
\end{split}
\label{eq:uncertainties}
\end{equation}
for every $\xi\in\mathbb R^{n_d}$ and almost every $(x,t)\in V\times T$. The bounds and the sign $s$ are considered known. For each well source region $V_j^*\subset V_{R_i}$, there exists a bounded Lipschitz neighborhood $U_j$ such that $V_j^*\subset U_j$, $\operatorname{dist}(V_j^*,\partial U_j)>0$, and $\overline{U_j}\subset V_{R_i}$. These neighborhoods are chosen so that $k/\eta\in W^{1,\infty}(U_j)$ and $\gamma_1$, $R^*$, and $\gamma_2(\cdot,t)$ belong to $H^2(U_j)$, with bounds uniform in time. 
\end{assumption}
\begin{assumption}\label{A4}:
The output regions in \eqref{eq:output} are pairwise disjoint, and each contains at least one well source region $V_k^*$. Every well source region is either contained in one output region or disjoint from all output regions. Moreover,
\begin{equation}
\begin{gathered}
 V_{u_i}\cap V_{u_j}=\emptyset\quad(i\ne j),\qquad
 V_{R_i}\cap V_{R_j}=\emptyset\quad(i\ne j),\\
 V_{u_i}\cap V_{R_j}=\emptyset,\qquad
 m=m_u+m_R\leq n,\\
 \min_{1\leq k\leq n}\operatorname{dist}\!\left(V_k^*,
 \partial V\cup\bigcup_{i=1}^{m_u}\partial V_{u_i}
 \cup\bigcup_{j=1}^{m_R}\partial V_{R_j}\right)>0.
\end{gathered}
\label{eq:region_separation}
\end{equation}
\end{assumption}
\begin{assumption}\label{A5}:
Let $p(x,t)=u(x,t)-|V|^{-1}\int_Vu(y,t)\,dV$, and let $A=-\beta^{-1}\nabla\cdot[(k/\eta)\nabla\,\cdot\,]$ denote the Neumann diffusion operator on $L^2(V)$. The initial pressure and its rate satisfy
\begin{equation}
\begin{split}
 p(0)\in D(A),\quad
 p_t(0)\in D(A).
\end{split}
\label{eq:initial_compatibility}
\end{equation}
The initial SR $R^0(x)=R(x,0)$ satisfies, for finite positive constants $L_{R^0}$, $L_{R^0,H^2}$, and $R_0^m$,
\begin{equation}
\begin{split}
 \|R^0\|_{L^\infty(V)}\leq L_{R^0},\qquad
 \|R^0\|_{H^2(U_j)}\leq L_{R^0,H^2},\\
 R^0(x)\geq R_0^m>0\quad\text{a.e. in }U_j.
\end{split}
 \label{eq:initial_SR_bounds}
\end{equation}
\end{assumption}

\begin{remark}
Assumptions~\ref{A1} and \ref{A2} are reasonable in practical control applications, as physical well capacities impose bounds on the applied fluxes, while reference trajectories can be selected to satisfy the desired operating conditions. This is consistent with reservoirs operating with finite production and injection volumes. Assumption~\ref{A3} considers positive and bounded material parameters, as expected from the physical properties of the reservoir, together with regularity conditions on the spatially varying coefficients. Assumption~\ref{A4} means that the pressure and SR output regions do not intersect, and that each selected region contains at least one well source region. These conditions can be satisfied by a suitable selection of the controlled regions, depending on the well distribution and reservoir geometry. Finally, Assumption~\ref{A5} imposes regularity and compatibility conditions on the initial pressure and SR fields.  
\end{remark}

\section{Output Feedback Tracking Control Design}
\label{sec:control}

Let us define the error variables, $\sigma_u \in \mathbb{R}^{m_u}$ and $\sigma_R \in \mathbb{R}^{m_R}$, as follows
\begin{equation}
\begin{split}
 \sigma_u(t) &= y_u(t)-r_u(t), \\
 \sigma_R(t) &= \frac{s}{\gamma_{1_0} R^*_0}\left[y_R(t)-r_R(t) \right],
\end{split}
\label{eq:error}
\end{equation}
where $\gamma_{1_0}>0,R^*_0>0$ are nominal values of $\gamma_1(x)$ and $R^*(x)$, respectively, that have to be selected. Note that the coefficient $\nicefrac{1}{\gamma_{1_0} R^*_0}$ was added in the second equation for units consistency.

Using the diffusion equation \eqref{eq:diff} and the SR system \eqref{eq:SR}, the error dynamics become
\begin{equation*}
\begin{split}
  \dot{\sigma}_{u_i}(t) &= -\frac{1}{\beta V_{u_i}}\int_{V_{u_i}} \nabla q(x,t) \, dV\\
  &\quad +\frac{1}{\beta V_{u_i}} \int_{V_{u_i}} \sum_{j=1}^{n} \mathcal{B}_j(x) Q_j(t) \, dV -\dot{r}_{u_i}(t), \quad i = 1,...,m_u,
\end{split}
\end{equation*}
\begin{equation*}
\begin{split}
  \dot{\sigma}_{R_i}(t) &= -\frac{1}{\gamma_{1_0} R^*_0 \beta V_{R_i}}\int_{V_{R_i}} \gamma_1(x) R(x,t) \nabla q(x,t) \, dV\\
  &\quad +\frac{1}{\gamma_{1_0} R^*_0 \beta V_{R_i}} \int_{V_{R_i}} \gamma_1(x) R(x,t) \sum_{j=1}^{n} \mathcal{B}_j(x) Q_j(t) \, dV \\
  &\quad - \frac{s}{\gamma_{1_0} R^*_0 V_{R_i}} \int_{V_{R_i}} \gamma_2(x,t) R(x,t) \left[R(x,t)-R^*(x) \right] \, dV \\
  &\quad -\frac{s}{\gamma_{1_0} R^*_0}\dot{r}_{R_i}(t), \quad i = 1,...,m_R.
\end{split}
\end{equation*}

In matrix form we have
\begin{equation}
  \dot{\sigma}(t) = \Psi(t)+ B(t)Q(t),
  \label{eq:errordyn}
\end{equation}
where $\sigma(t)=[\sigma_u(t)^T,\sigma_R(t)^T]^T$, $\Psi(t) \in \mathbb{R}^{m}$ is defined as $\Psi(t) = \begin{bmatrix}
        \psi_1(t) \in \mathbb{R}^{m_u} \\ \psi_2(t) \in \mathbb{R}^{m_R}
    \end{bmatrix}$ with
    
\begin{equation}
\begin{split}
    \psi_1(t) &= \begin{bmatrix}
        -\frac{1}{\beta V_{u_i}}\int_{V_{u_i}} \nabla q(x,t) \, dV - \dot{r}_{u_i}(t)
    \end{bmatrix}_{i=1}^{m_u}, \\
    \psi_2(t) &= \begin{bmatrix}
        -\frac{1}{\gamma_{1_0} R^*_0 \beta V_{R_i}}\int_{V_{R_i}} \gamma_1(x) R(x,t) \nabla q(x,t) \, dV \\
        -\frac{s}{\gamma_{1_0} R^*_0 V_{R_i}} \int_{V_{R_i}} \gamma_2(x,t) \left[R(x,t)\right]^2 \, dV \\
        + \frac{s}{\gamma_{1_0} R^*_0 V_{R_i}} \int_{V_{R_i}} \gamma_2(x,t) R(x,t)R^*(x) \, dV \\
        - \frac{s}{\gamma_{1_0} R^*_0}\dot{r}_{R_i}(t)
    \end{bmatrix}_{i=1}^{m_R},
\end{split}
\label{eq:Psi}
\end{equation}
and $B(t) \in \mathbb{R}^{m \times n}$ is defined as $B(t)=[B_u,B_R(t)]^T$ with $B_u=[b_{ij}^u] \in \mathbb{R}^{m_u \times n}$, $B_R(t)=[b_{ij}^R(t)] \in \mathbb{R}^{m_R \times n}$ defined as
{\small
\begin{equation}
\begin{split}
  b_{ij}^u &= \left\{ \begin{array}{c}
  \frac{1}{\beta V_{u_i}} \\ 
  0
  \end{array} \right. 
  \begin{array}{l}
  \textup{if} \quad V_j^* \subset V_{u_i} \\ 
  \textup{if} \quad V_j^* \not\subset V_{u_i}
  \end{array} ,
  \begin{array}{l}
  i=1,...,m_u \\ 
  j=1,...,n
  \end{array} , \\
  b_{ij}^R(t) &= \left\{ \begin{array}{c}
  {\scriptstyle \frac{1}{\gamma_{1_0} R^*_0 \beta V_{R_i} V_j^*} \int_{V_j^*} \gamma_1(x) R(x,t) \, dV} \\ 
  0
  \end{array} \right. 
  \begin{array}{l}
  \textup{if} \quad V_j^* \subset V_{R_i} \\ 
  \textup{if} \quad V_j^* \not\subset V_{R_i}
  \end{array} , \\
  &\hspace{170pt}
  \begin{array}{l}
  i=1,...,m_R \\ 
  j=1,...,n
  \end{array} .
\end{split}
  \label{eq:B1}
\end{equation}}
where the definition of $\mathcal{B}_i(x)$ in \eqref{eq:B} has been used. Notice that $B(t)$ does not depend explicitly on $s$, due to the definition of $\sigma_R$ in \eqref{eq:error}.

The uncertain control coefficient is defined from the product of the actual and nominal input matrices as
\begin{equation}
  G(t):=B(t)B_0^{+}=\mathbb{I}_m+\Delta B(t),
  \label{eq:Bbounds}
\end{equation}
where $\Delta B(t)\in\mathbb{R}^{m\times m}$ and $B_0\in\mathbb{R}^{m\times n}$ is the nominal (known) control coefficient. The matrix $B_0^{+} \in \mathbb{R}^{n \times m}$ is the Moore--Penrose right pseudoinverse of $B_0$, which exists due to Assumption~\ref{A4}. The matrix $B_0$ is chosen as $B_0=[B_{u_0},B_{R_0}]^T$ with $B_{u_0}=[b_{ij}^{u_0}] \in \mathbb{R}^{m_u \times n}$, $B_{R_0}=[b_{ij}^{R_0}] \in \mathbb{R}^{m_R \times n}$ defined as
\begin{equation}
\begin{split}
  b_{ij}^{u_0} &= \left\{ \begin{array}{c}
  \frac{1}{\beta_0 V_{u_i}} \\ 
  0
  \end{array} \right. 
  \begin{array}{l}
  \textup{if} \quad V_j^* \subset V_{u_i} \\ 
  \textup{if} \quad V_j^* \not\subset V_{u_i}
  \end{array} ,
  \begin{array}{l}
  i=1,...,m_u \\ 
  j=1,...,n
  \end{array} , \\
  b_{ij}^{R_0} &= \left\{ \begin{array}{c}
  \frac{1}{\beta_0 V_{R_i}} \\ 
  0
  \end{array} \right. 
  \begin{array}{l}
  \textup{if} \quad V_j^* \subset V_{R_i} \\ 
  \textup{if} \quad V_j^* \not\subset V_{R_i}
  \end{array} ,
  \begin{array}{l}
  i=1,...,m_R \\ 
  j=1,...,n
  \end{array} .
\end{split}
  \label{eq:B01}
\end{equation}
\noindent where $\beta_0>0$ is a nominal value of $\beta$ that has to be selected. Notice that all the nominal matrices are constant and, as such, they require minimum measurements on \eqref{eq:errordyn}, \textit{i.e.}, we do not need to measure the terms $\int_{V_j^*} \gamma_1(x)R(x,t) \, dV$, $j=1,...,n$ in \eqref{eq:B1}.  


Let us choose the control $Q(t)$ as
\begin{equation}
\begin{split}
  Q(t) &= B_0^{+} \left[-K_1 \Sabs{\sigma(t)}^{\frac{1}{1-l}} + \nu(t) \right], \\
  \dot{\nu}(t) &= -K_2 \Sabs{\sigma(t)}^{\frac{1+l}{1-l}},
\end{split}
\label{eq:Q}
\end{equation}
where $K_1 \in \mathbb{R}^{m \times m}$, $K_2 \in \mathbb{R}^{m \times m}$ are matrices to be designed. The function $\lceil \sigma \rfloor^{\gamma}:=|\sigma|^{\gamma}\sign(\sigma)$ is applied element-wise and is determined for any $\gamma\in \Re_{\geq 0}$. Such nonlinear control has different characteristics depending on the value of $l \in [-1,0]$ \citep{b:Mathey-Moreno-2024}. It has a discontinuous integral term when $l=-1$, and it is known as the MIMO Super-Twisting. When $l \in (-1,0]$ the control function is continuous and turns into a linear integral control when $l=0$. The control signal generated is always continuous despite having a possible discontinuous integral term. In this case, the solutions of the closed-loop systems \eqref{eq:diff}, \eqref{eq:errordyn} with control \eqref{eq:Q} are understood in the sense of Filippov \citep{b:filippov,b:Orlov-2020}. Note also that the controller is designed with minimum information about the system \eqref{eq:errordyn}, \textit{i.e.}, with only the measurement of $\sigma(t)$ and the knowledge of the nominal matrix $B_0$. 

A physical interpretation of the control law for SR mitigation is as follows: control \eqref{eq:Q} adjusts the well fluxes in response to deviations in the region-averaged pressure and SR. The super-twisting structure ensures robustness against model uncertainties and spatial heterogeneities.

Let us define a new variable $\sigma_I(t) = \nu(t) + \bar{\Psi}(t)$ with
\begin{equation}
  \bar{\Psi}(t):=G(t)^{-1} \Psi(t).
  \label{eq:Psibar}
\end{equation}
Then, the closed-loop system \eqref{eq:errordyn}, \eqref{eq:Q} can be written as
\begin{equation}
\begin{split}
  \dot{\sigma} &= G(t)\left[-K_1 \Sabs{\sigma}^{\frac{1}{1-l}} + \sigma_I \right], \\
  \dot{\sigma}_I &= -K_2 \Sabs{\sigma}^{\frac{1+l}{1-l}} + \dot{\bar{\Psi}}(t).
\end{split}
\label{eq:closed}
\end{equation}
The following tracking result then holds for system \eqref{eq:closed}.

\begin{theorem}
Consider system \eqref{eq:errordyn} under Assumptions \ref{A1}--\ref{A5}, driven by control law \eqref{eq:Q}. Then there exist nominal values of $\beta_0$, $\gamma_{1_0}$ and $R^*_0$ such that 
\begin{align}
  &\norm{G(t)-\mathbb{I}_{m}} \leq \delta_B < 1, \label{eq:DeltaBbound}\\
  &\norm{\dot{\bar{\Psi}}(t)} \leq \delta <\infty, \label{eq:Psi2bound}
\end{align}
If $\delta_B$ satisfies the admissible smallness condition of \citep{b:Mathey-Moreno-2024}, there exist matrices $K_1,K_2$ defined as
\begin{equation}
\begin{split}
  K_1&=L\bar{K}_1, \quad K_2=L^2\bar{K}_2,
\end{split}
\label{eq:ks}
\end{equation}
where $L>0$ is sufficiently large, and $\bar{K}_1,\bar{K}_2$ are positive diagonal matrices selected according to \citep{b:Mathey-Moreno-2024}, such that the origin of system \eqref{eq:closed} is:
\begin{enumerate}
  \item finite-time stable for $l=-1$.
  \item input-to-state stable with respect to $\dot{\bar{\Psi}}(t)$ for $l\in(-1,0)$, and finite-time stable if $\dot{\bar{\Psi}}(t)=0$.
  \item input-to-state stable with respect to $\dot{\bar{\Psi}}(t)$ for $l=0$, and exponentially stable if $\dot{\bar{\Psi}}(t)=0$.
\end{enumerate}
\label{th:1}
\end{theorem} 

\begin{proof}
The proof is shown in Appendix~\ref{app:control}.
\end{proof}

\begin{remark}
The control \eqref{eq:Q} generates a continuous control signal,
but does not automatically respect prescribed actuator limits for
arbitrary initial conditions. Therefore, the theorem is conditional
on Assumption~\ref{A1}. Saturated and conditioned super-twisting
implementations are discussed in
\citep{b:SeeberHornSaturated2019,b:SeeberReichhartingerConditioned2020}
and could provide a basis for extending the present stability
analysis to flux saturation in future work.
\end{remark}

\begin{remark}:
Theorem \ref{th:1} can be also obtained for Dirichlet BCs in the diffusion equation \eqref{eq:diff}, \textit{i.e.}, $u(x,t)=0$ for all $x \in \partial V$ (see also Appendix~\ref{app:control_bound_diff}). 
\end{remark}

\subsection*{Demand and input constraints}

Following \citep{b:Gutierrez-Stefanou-2024}, we will consider a new feature where an additional number of flux restrictions, $n_r$, with $n_r+m \leq n$, over the inputs $Q(t)$ of system \eqref{eq:diff} is needed. In other words, we will impose the weighted sum of the injection rates of the inputs to be equal to time-varying functions $D(t)$. 

The condition imposed over the control input, $Q(t)$, is
\begin{equation}
    W Q(t) = D(t),
    \label{eq:restriction}
\end{equation}
where $W \in \mathbb{R}^{n_r \times n}$ is a full rank matrix whose elements represent the weighted participation of the input fluxes for ensuring the demand/production $D(t) \in \mathbb{R}^{n_r}$. In order to follow this constraint, the control input is modified as follows
\begin{equation}
\begin{split}
   Q(t) &= \overline{W} \left(B_0 \overline{W} \right)^{+} \left[-K_1 \Sabs{\sigma(t)}^{\frac{1}{1-l}} + \nu(t) \right] \\
   &\quad + W^T \left(W W^T \right)^{-1}D(t), \\
  \dot{\nu}(t) &= -K_2 \Sabs{\sigma(t)}^{\frac{1+l}{1-l}},
\end{split}
   \label{eq:Qr}
\end{equation}
where $\sigma(t)$ is the original error vector, and the columns of $\overline{W} \in \mathbb{R}^{n\times(n-n_r)}$ form an orthonormal basis of the null space of $W$. Note that if we replace \eqref{eq:Qr} in \eqref{eq:restriction}, the demand over the controlled injection points will be satisfied exactly at any time $t$. 

Let us define the variable $\sigma_I(t) = \nu(t) + \bar{\Psi}_W(t)$ with
\begin{equation}
\begin{split}
  \bar{\Psi}_W(t) &:= G_W(t)^{-1} \Psi_W(t), \\
  G_W(t) &= B(t)\overline{W}\left(B_0\overline{W}\right)^+,\\
  \Psi_W(t) &= \Psi(t)+B(t)W^T(WW^T)^{-1}D(t).
\end{split}
  \label{eq:Psibar2}
\end{equation}
Then, the closed-loop system \eqref{eq:errordyn}, \eqref{eq:Qr} can be written as
\begin{equation}
\begin{split}
  \dot{\sigma} &= G_W(t)\left[-K_1 \Sabs{\sigma}^{\frac{1}{1-l}} + \sigma_I \right], \\
  \dot{\sigma}_I &= -K_2 \Sabs{\sigma}^{\frac{1+l}{1-l}} + \dot{\bar{\Psi}}_W(t).
\end{split}
\label{eq:closed2}
\end{equation}
The conditions on the demand allocation, $W$, and the proof of its tracking properties are given below.

\begin{assumption}\label{A6}
The matrix $W$ has full row rank and $B_0\overline W$ has full row rank. Moreover, the matrix $W$ is chosen such that condition \eqref{eq:WpositiveNew} holds for all $h_j\in[h_j^-,h_j^+]$, where $h_j$ is the average of $\gamma_1 R$ over the well region $V_j^*$, and $h_j^-$ and $h_j^+$ are its positive lower and upper bounds, respectively, defined in Appendix~\ref{app:demand}. The demand $D(t)$ is absolutely continuous with bounded derivative, $\norm{\dot D(t)}\leq L_D$ for all $t\in T$, and is compatible with the prescribed well capacities.
\end{assumption}

\begin{theorem}\label{th:demand}
Consider system \eqref{eq:diff}--\eqref{eq:SR} under Assumptions~\ref{A1}--\ref{A6}, driven by the constrained control law \eqref{eq:Qr}. Then, there exist nominal values of $\beta_0$, $\gamma_{1_0}$ and $R^*_0$ such that 
\begin{align}
  &\norm{G_W(t)-\mathbb I_m}\leq\delta_{B,W}<1, \label{eq:DeltaBbound2}\\
  &\norm{\dot{\bar{\Psi}}_W(t)} \leq\delta_W<\infty. \label{eq:Psi2bound2}
\end{align}
For gains selected as in \eqref{eq:ks} under the admissible uncertainty conditions of \citep{b:Mathey-Moreno-2024}, the origin of system \eqref{eq:closed2} is:
\begin{enumerate}
  \item finite-time stable for $l=-1$.
  \item input-to-state stable with respect to $\dot{\bar{\Psi}}_W(t)$ for $l\in(-1,0)$, and finite-time stable if $\dot{\bar{\Psi}}_W(t)=0$.
  \item input-to-state stable with respect to $\dot{\bar{\Psi}}_W(t)$ for $l=0$, and exponentially stable if $\dot{\bar{\Psi}}_W(t)=0$.
\end{enumerate}
\end{theorem}

\begin{proof}
The proof is given in Appendix~\ref{app:demand}.
\end{proof}

As an additional verification step, both theorems \ref{th:1} and \ref{th:demand} were verified in Lean~4.35.0-rc4 and machine-checked with Mathlib v4.35.0-rc4 (\url{https://github.com/ERC-INJECT/reservoir-seismicity-control}).

\section{Mitigation of Induced Seismicity in Groningen Reservoir}
\label{sec:simulations}

The Groningen gas field, located in the northeastern Netherlands, is one of the largest natural gas fields in both Europe and the world, with an estimated 2,900 billion cubic meters of recoverable gas. However, gas extraction in Groningen has triggered earthquakes since 1991, causing cracking and other structural damage to buildings, and raising concern among residents. \citep{Bourne2014,Candela2019Groningen} In June 2023, the Dutch government announced that gas extraction would end by October 1, 2023, leaving about 470 billion cubic meters of gas still in the field \citep{b:Groningen1,b:Groningen2}. 
\begin{figure}[ht!]
  \centering
  \includegraphics[width=7.0cm,keepaspectratio]{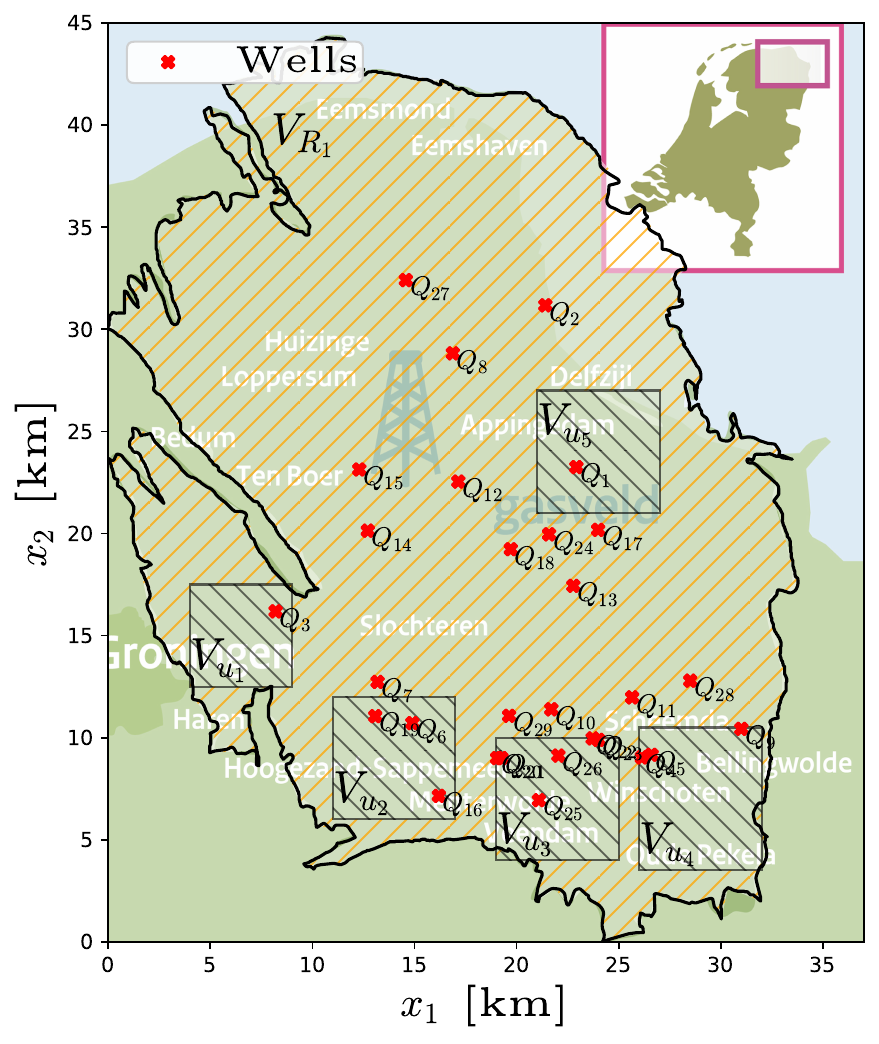}
  \caption{Groningen gas reservoir with regions for controlling pressure and SR based on the locations of major towns in the area. Background image adapted from \url{https://zoek.officielebekendmakingen.nl/stcrt-2017-28922.html}.}
  \label{fig:Groningen}
\end{figure}

Gas extraction in Groningen causes pressure depletion, reservoir compaction and poroelastic stress changes that can promote fault reactivation and seismicity \citep{Bourne2014,DempseySuckale2017,Candela2019Groningen,b:SMITH2022117697,b:doi.org/10.1029/2023GL105455}. The hydro-mechanical behaviour of the reservoir due to fluid injection/extraction can be described by Biot's theory \citep{b:Biot-1941}, which couples fluid diffusion and rock deformation. However, when fluid injection rates are slow and rock volumetric strain is negligible, the diffusion of the fluid in the host rock due to fluid injections/extractions of the wells is further simplified to the diffusion equation \eqref{eq:diff} \citep{Zienkiewicz1980}. In Groningen, $n=29$ wells are located along the reservoir (see Fig. \ref{fig:Groningen}) and undrained boundary conditions are considered at the boundary, \textit{i.e.}, $q(x,t) \cdot \hat{e}=0$ at $\partial V$ (see \citep{b:SMITH2022117697,b:doi.org/10.1029/2023GL105455,b:10.1785/0220230179,b:https://doi.org/10.1029/2024GL110139,b:https://doi.org/10.1029/2020JB020013} for more details of the BCs in Groningen). Here, each modeled well represents a production cluster in the real reservoir comprising several individual wells.

Following \citep{Segall2015,Dieterich1994,b:SMITH2022117697,b:doi.org/10.1029/2023GL105455,b:10.1785/0220230179,b:https://doi.org/10.1029/2024GL110139,b:https://doi.org/10.1029/2020JB020013,b:kim2023,Wassing2021,https://doi.org/10.1029/2018WR023587}, system \eqref{eq:SR}, with $s=1$, corresponds to a Dieterich-type SR model with positive pressure-rate coupling. In other words, it expresses the number of seismic events per unit time in a given region. Dieterich-type SR models have been used for Groningen \citep{b:SMITH2022117697,b:doi.org/10.1029/2023GL105455}, Otaniemi \citep{b:kim2023}, and Pohang \citep{Wassing2021}, among other reservoirs. We introduced $s=-1$ to account for cases where the increase of seismicity is due to fluid extraction, which is the assumption adopted for Groningen in this work.  

In \citep{Segall2015,Dieterich1994}, the SR system is defined by $R^n_t = \frac{R^n}{t_a}\left(\frac{\dot{\tau}}{\dot{\tau}_0}-R^n \right)$, where $\dot{\tau}$ is the Coulomb stressing rate, $\dot{\tau}_0$ is the background stressing rate, and $t_a$ is a characteristic decay time. We adopt the effective relation $\dot{\tau} = \dot{\tau}_0 + f s u_t$, where $f>0$ represents the effective pressure-to-stress coupling. For Groningen, we assume \(s=-1\). This sign does not represent the direct effect of pore-pressure decrease on a fault, which increases effective normal stress and tends to stabilize it \citep{Candela2018}. Instead, pressure depletion causes compaction and poroelastic stress redistribution that can promote slip \citep{Segall1989,Bourne2014,Candela2019Groningen,b:SMITH2022117697}. Moreover, many earthquakes in Groningen occur in the overburden above the reservoir rather than within the reservoir itself \citep{Smith2020}. The simplified relation approximates this net stressing response and neglects fault geometry and the detailed compaction process. With this relation, the normalized SR dynamics becomes $R^n_t = \frac{R^n}{t_a}\left(\frac{f}{\dot{\tau}_0} s u_t + 1 - R^n \right)$. Setting $R(x,t) = R^n(x,t) R^*(x)$, $\gamma_1(x) = \frac{f}{\dot{\tau}_0 t_a}$ and $\gamma_2(x,t) = \frac{1}{t_a R^*}$ we recover \eqref{eq:SR}, which allows a more general description of the process. Note that \eqref{eq:SR} is defined point-wise, differently from \citep{b:Gutierrez-Stefanou-2024}, where the SR was defined region-wise. Therefore, $R(x,t)$ denotes the point-wise SR density in events/(year$\cdot$km$^3$), $u_t(x,t)$ is the input of this system and denotes the partial derivative of $u(x,t)$ with respect to time. The background SR is given by $R^*(x)$, which represents the intrinsic SR of the region in the absence of fluid injection or extraction. The parameter $\gamma_1=f/(t_a\dot\tau_0)$ has units of MPa$^{-1}$ and represents the sensitivity of the SR to changes in pressure rate, including the effective stress coupling and the characteristic relaxation time. In the rate-and-state formulation, $t_a=a\sigma_{\mathrm{eff}}/\dot\tau_0$, where $a>0$ is a dimensionless rate-and-state friction parameter and $\sigma_{\mathrm{eff}}>0$ is the effective normal stress \citep{b:doi.org/10.1029/2023GL105455}. Therefore, $\gamma_1=f/(a\sigma_{\mathrm{eff}})$. The parameter $\gamma_2=1/(t_aR^*)$ has units of km$^3$/event and represents a relaxation coefficient accounting for the background SR density. In the normalized equation, the corresponding relaxation coefficient is $1/t_a$. However, in the present Groningen numerical validation, $\gamma_1,\gamma_2$ are also considered as fitting parameters that can represent various physical mechanisms, provided that the final model captures the observed seismicity of the reservoir.

\begin{figure}[ht!]
  \centering
  \includegraphics[width=6.5cm,keepaspectratio]{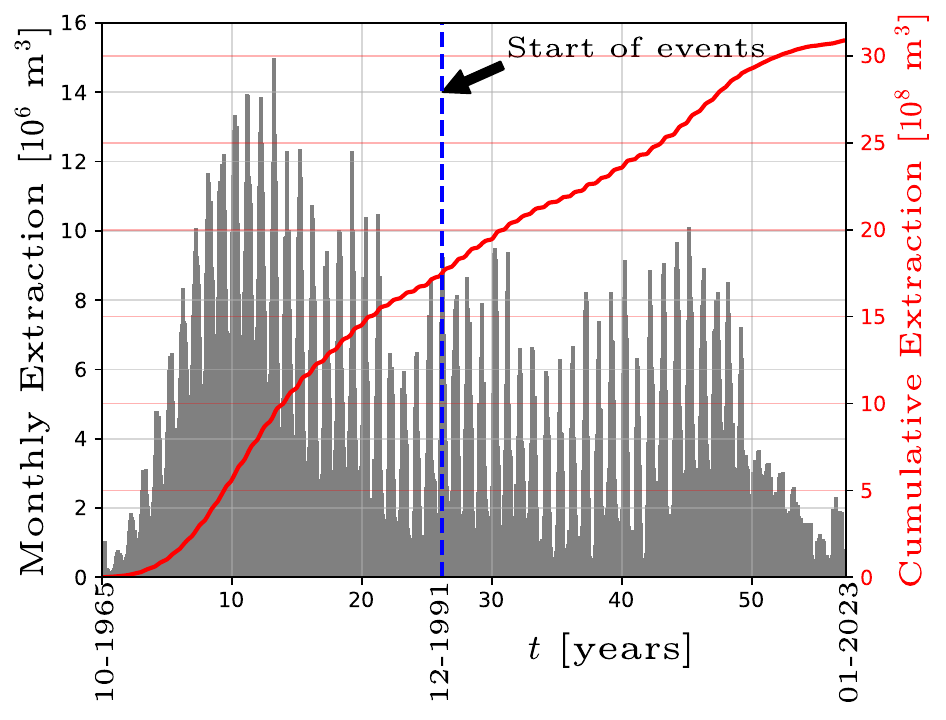}
  \caption{Monthly and cumulative extraction of gas in Groningen.}
  \label{fig:extraction}
\end{figure}

\begin{figure}[ht!]
  \centering
  \includegraphics[width=13cm,keepaspectratio]{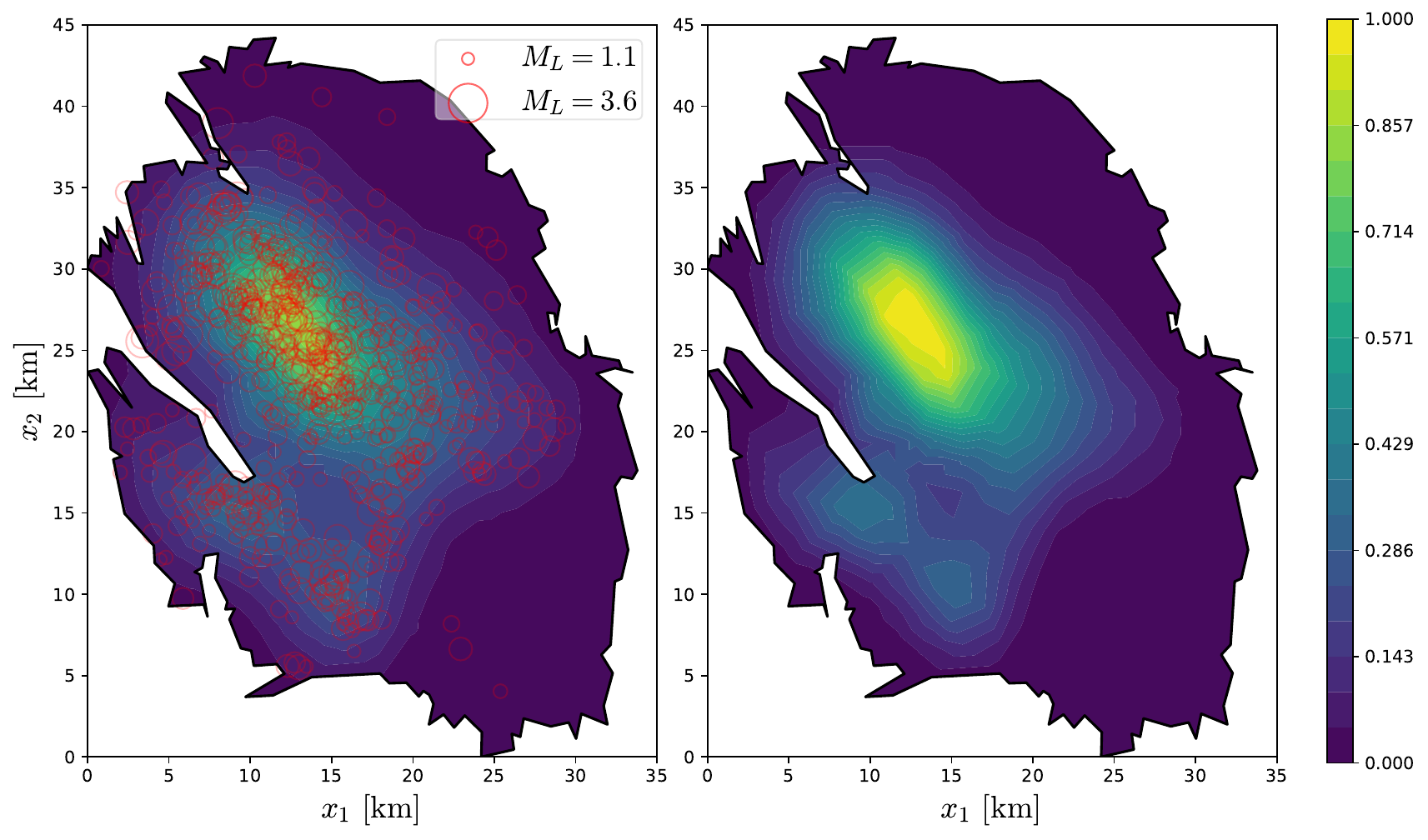}
  \caption{Normalized spatial density map of the SR in Groningen representing the 690 events selected between 12-1991 and 01-2023. The left image shows the magnitude and location of the observed events, from which their spatial density is determined. The right image shows the simulated spatial density. Both are normalized by the maximum spatial density.}
  \label{fig:density}
\end{figure}

\begin{figure}[ht!]
  \centering
  \includegraphics[width=6.5cm,keepaspectratio]{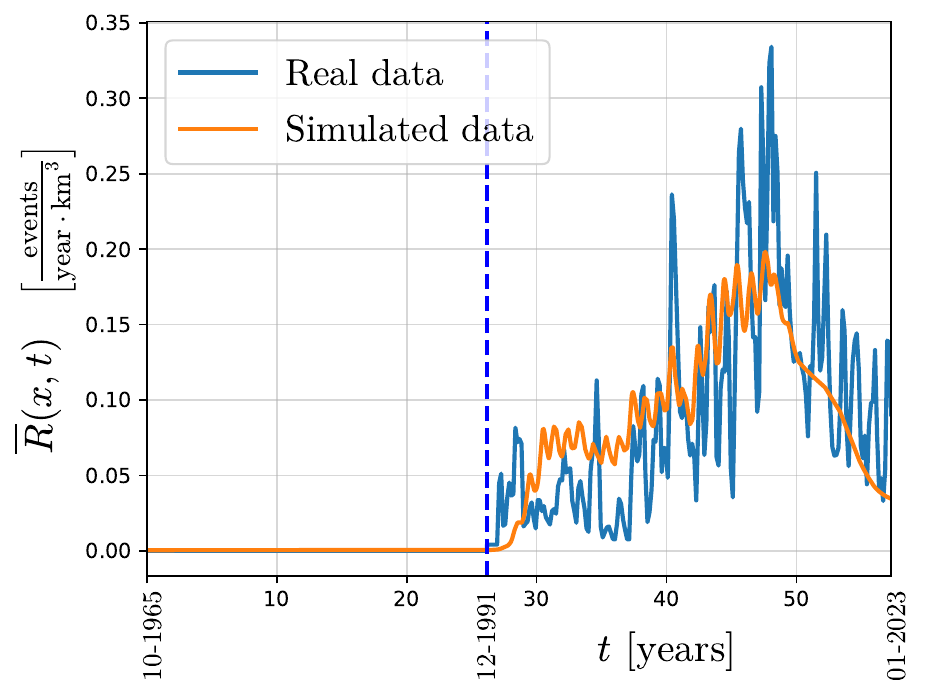}
  \includegraphics[width=6.5cm,keepaspectratio]{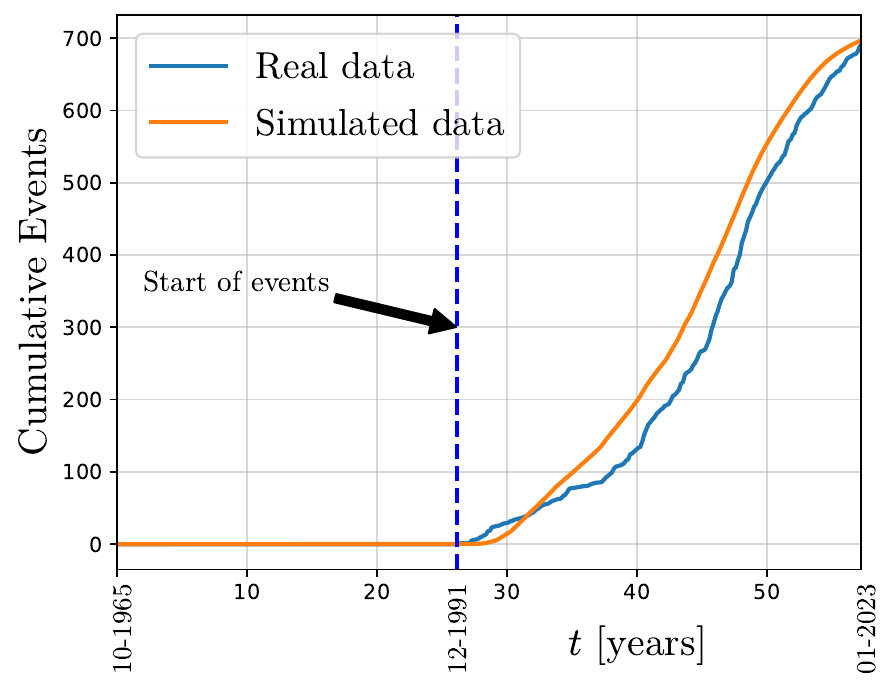}
  \caption{Average SR density over the reservoir and cumulative number of seismic events in Groningen. Real data (blue line) and simulated data (orange line). The variations of the simulated SR reflect the time-varying historical monthly extraction rates.}
  \label{fig:validation}
\end{figure}

In the absence of pressure variations, $u_t(x,t)=0$ and, therefore, $R(x,t)\rightarrow R^*(x)$. In this case, the SR of the region reduces to the background one. Depending on the sign of \(s\), fluid injection or extraction may result in \(su_t(x,t)>0\), which promotes an increase in seismicity above its background value under the adopted model. This is demonstrated from real data and modelling of the reservoir (see \citep{b:Groningen1,b:Groningen2,b:SMITH2022117697,b:doi.org/10.1029/2023GL105455,b:10.1785/0220230179,b:acosta_2023_8329298,b:https://doi.org/10.1029/2024GL110139,b:https://doi.org/10.1029/2020JB020013}), between 10-1965 to 01-2023. Fig. \ref{fig:extraction} shows the total gas extraction history ($-\sum_{i=1}^{29} Q_i(t)$) in the whole reservoir. The spatial distribution of all the events during the study period is shown in the left side of Fig. \ref{fig:density}. Fig. \ref{fig:validation} (blue line) shows the average SR density over the whole reservoir, $\bar{R}(t)=|V|^{-1}\int_V R(x,t)\,dV$, and the cumulative number of events, $\int_0^t\int_V R(x,s)\,dV\,ds$. The selected Royal Netherlands Meteorological Institute (KNMI) earthquake catalog contains 690 events from December 1991 to January 2023, with a completeness magnitude $M_c=1.1$. The catalog and the gas extraction history used in the simulations were taken from the data associated with \citep{KNMI2023,b:doi.org/10.1029/2023GL105455,b:acosta_2023_8329298}. 

Gas extraction has caused earthquakes in the Groningen reservoir, leading to its closure. Here, we use control \eqref{eq:Qr} to reduce operation-induced seismicity while imposing weighted production-demand constraints based on historical extraction, starting from 12-1991. For that purpose, we will first select parameters and coefficients for systems \eqref{eq:diff}--\eqref{eq:SR} to fit and validate the model against real reservoir data.

\subsection{Model setup and validation}

We start by selecting the hydraulic parameters of system \eqref{eq:diff} according to \citep{b:SMITH2022117697,b:doi.org/10.1029/2023GL105455,b:10.1785/0220230179,b:https://doi.org/10.1029/2024GL110139,b:https://doi.org/10.1029/2020JB020013}: the hydraulic diffusivity is $c_{hy} = 4.4 \times 10^{-2}$ [km$^2$/hr], and the mixture compressibility is $\beta = 5.7 \times 10^{-4}$ [MPa$^{-1}$]. According to those values, $c_{hy}(x) = \nicefrac{k(x)}{\beta \eta(x)}$ is constant. The parameter $c_{hy}$ represents the hydraulic diffusivity, which characterizes the ability of a porous medium to diffuse fluid in response to pressure variations. A higher value of $c_{hy}$ indicates that fluid can propagate faster through the medium, leading to faster pressure equilibration. We then verify that the average pressure over the reservoir matches the one shown in \citep[Figure 1]{b:doi.org/10.1029/2023GL105455}. For that purpose, we depth-average equation \eqref{eq:diff} and we integrate the resulting partial differential equation in time and space using finite elements as explained in Appendix \ref{app:sim}. The historical well rates $Q_i(t)$ were taken from the production data associated with \citep{b:doi.org/10.1029/2023GL105455,b:acosta_2023_8329298}.

\begin{figure}[ht!]
  \centering
  \includegraphics[width=6.5cm,keepaspectratio]{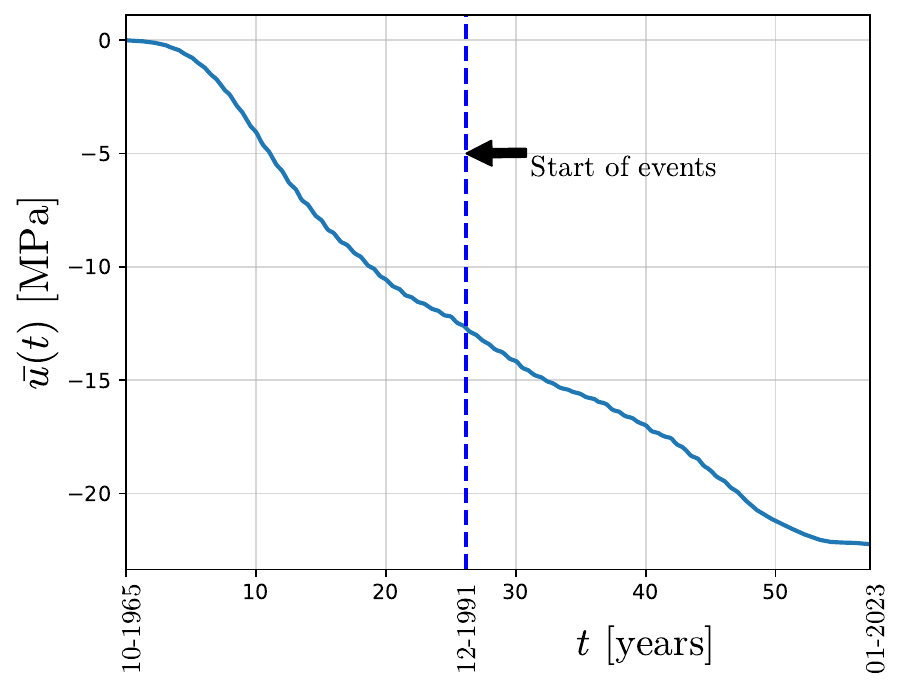}
  \caption{Average pore-pressure change over the reservoir relative to its initial state (see also \citep[Figure 1]{b:doi.org/10.1029/2023GL105455}).}
  \label{fig:uvalidation}
\end{figure}

The average fluid pressure over the reservoir is shown in Fig. \ref{fig:uvalidation} where similar results were obtained to the ones reported in \citep[Figure 1]{b:doi.org/10.1029/2023GL105455} using more detailed models for the Groningen reservoir.

For the validation of system \eqref{eq:SR}, an optimization algorithm was implemented to select $\gamma_1(x),\gamma_2(x,t)$ based on the normalized spatial density of the real SR data shown in Fig. \ref{fig:density} (left side). The normalized simulated density, $d(x)$, is shown in Fig \ref{fig:density} (right side). The optimized parameters are $\gamma_1(x)=\gamma_1^M d(x) > 0$, where the maximum fitted pressure sensitivity is $\gamma_1^M = 7.79$ [MPa$^{-1}$], $\gamma_2(x,t)=\gamma_2 = 4.26$ [km$^3$/event] is the relaxation coefficient of the unnormalized SR model, and the background SR density is $R^*(x)=R^* \approx 4.11 \times 10^{-4}$ [events/year/km$^3$], which corresponds to an expected value of approximately $1$ event per decade over the whole reservoir volume. The fitted spatial distribution of $\gamma_1(x)$ remains strictly positive throughout the numerical domain, consistently with Assumption~\ref{A3}. The chosen value of $R^*$ is a background parameter of the numerical model, not an independently measured natural seismicity rate. The average SR density and cumulative events obtained from the model reproduce the main trends of the observed data (Fig. \ref{fig:validation}). These results support its use as a reduced numerical model, rather than an independent validation of a fully three-dimensional Groningen reservoir. The variations of the simulated SR are due to the historical time-varying well rates, which act through $u_t$ in \eqref{eq:SR}. 

Two academic control scenarios are explored starting from December 1991, when seismic events began to be recorded. In the first scenario, control law \eqref{eq:Qr} is applied to regulate the average pressure over five regions of the reservoir, denoted by $V_{u_i}$, $i=1,\ldots,5$, while controlling the average seismicity rate in the complementary region $V_{R_1}$. Simultaneously, the controller must satisfy production-demand constraints with the total demand (see Fig.~\ref{fig:extraction}) distributed among the wells according to the imposed participation matrix (see \eqref{eq:restriction}). The selection of these five regions is based on the locations of major towns in the Groningen area and is illustrated in Fig. \ref{fig:Groningen}. In the second scenario, we will introduce an additional constraint on the input $Q(t)$ by incorporating CO$_2$ injection. This scenario is a conceptual test of the additional actuation provided by injection wells.

\subsection{Scenario 1: Gas extraction}

The control \eqref{eq:Qr} was implemented with the demand $D(t)=-f(t)$, where $f(t)$ is the extraction history shown in Fig. \ref{fig:extraction}, and the negative sign represents extraction. The weight matrix $W \in \mathbb{R}^{1 \times 29}$ was obtained by minimizing $\|\overline{W}(B_0\overline{W})^{+}\|$, subject to its entries lying in $[0.8,1.2]$ and $B_0\overline{W}$ having full row rank. This reduces the nominal control effort while satisfying the prescribed weighted demand. Since the weights need not be equal to one, $WQ=D$ does not necessarily prescribe the physical net extraction $-\sum_{i=1}^{29}Q_i(t)$. Under no-flux boundary conditions, $\beta|V|\dot{\bar{u}}(t)=\sum_{i=1}^{29}Q_i(t)$, where $\bar{u}(t)=|V|^{-1}\int_Vu(x,t)\,dV$. Thus, a prescribed net extraction fixes the mean pressure change rate, but does not determine the spatial pressure or nonlinear SR distribution. The tracking error was defined as in \eqref{eq:output} and \eqref{eq:error}, with $R^*_0=R^*$. For simplicity, the nominal parameter $\gamma_{1_0}$ was selected as the midpoint of the known bounds on $\gamma_1(x)$, giving $\gamma_{1_0}=(\gamma_1^m+\gamma_1^M)/2=3.90$ MPa$^{-1}$.

The pressure references were chosen ad-hoc to guide the outputs to an average pressure after 15 [years] using a sigmoid profile, while the SR reference was maintained at $R^*$. Therefore, the background SR, $R^*$, was set as a target in this scenario. In this model, the target corresponds to the predicted background seismicity, rather than a site-specific operational threshold. Higher positive references are possible but the small background SR value is more challenging for testing the control, and therefore, more appealing. Setting a smaller target for the SR close to zero could be also considered in some scenarios as well. However, the SR would inevitably return to the background SR after the end of our interventions. In this sense, reducing the SR below the background one, $R^*$, is less interesting.

The nominal matrix $B_0$ was selected as in \eqref{eq:B01}, with $\beta_0=0.8 \beta$, in accordance with condition \eqref{eq:constants}. Then, the control gains were designed according to \eqref{eq:ks}, resulting in $l=-1$, $K_1=\mathrm{diag}(47.43,47.43,47.43,47.43,47.43,6.71) \times 10^{-4}$, and $K_2=\mathrm{diag}(110,110,110,110,110,2.2) \times 10^{-7}$. The parameter $L$ in \eqref{eq:ks} sets the common gain scale and affects the closed-loop response and control effort once the admissible gain and uncertainty conditions hold. In the simulations it was selected to obtain the desired convergence and overshoot. Notice that the choice of the gains can be further optimized (see \textit{e.g.}, gain scheduling with reinforcement learning \citep{b:https://doi.org/10.1002/nag.3923}) but this goes beyond the target of this work.

\begin{figure}[ht!]
  \centering
  \hspace{-8pt}\includegraphics[width=6.8cm,keepaspectratio]{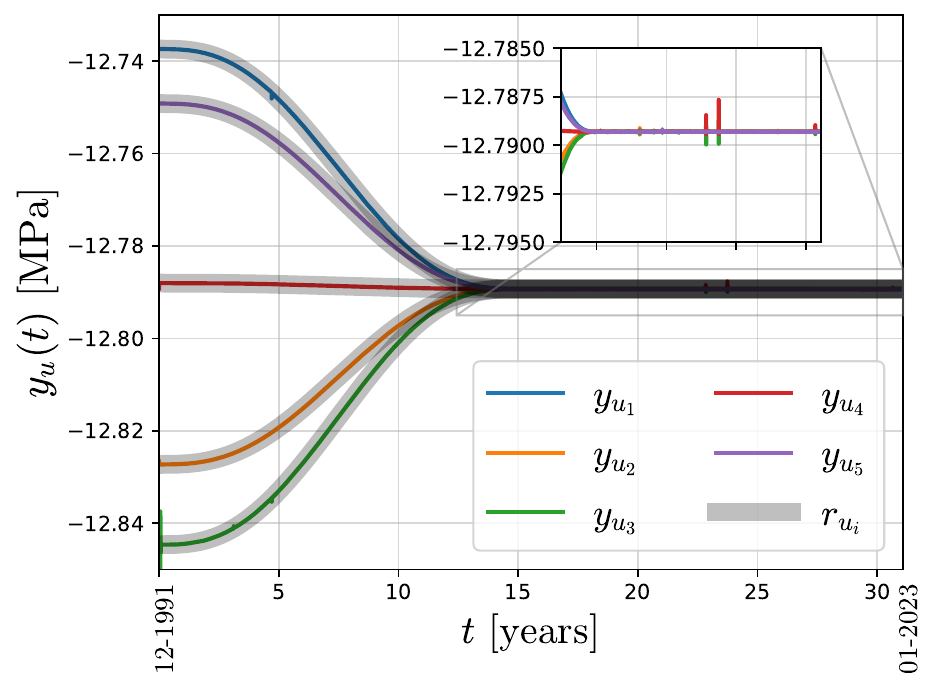}
  \includegraphics[width=6.5cm,keepaspectratio]{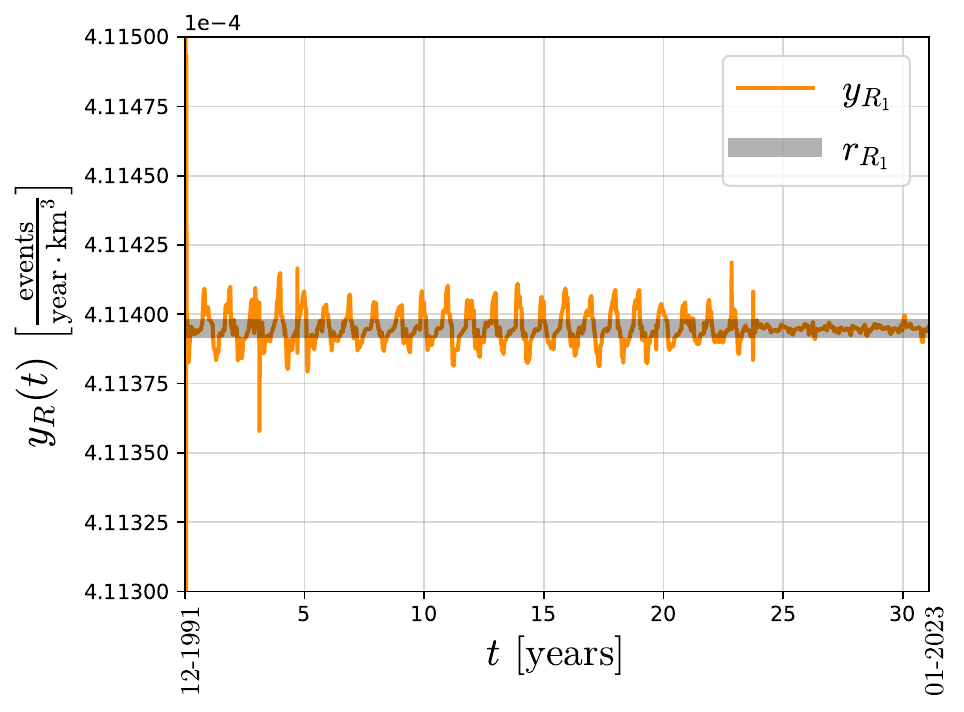}
  \caption{Pressure (top) and SR (bottom) outputs in Scenario 1.}
  \label{fig:output1}
\end{figure}

\begin{figure}[ht!]
  \centering
  \includegraphics[width=6.5cm,keepaspectratio]{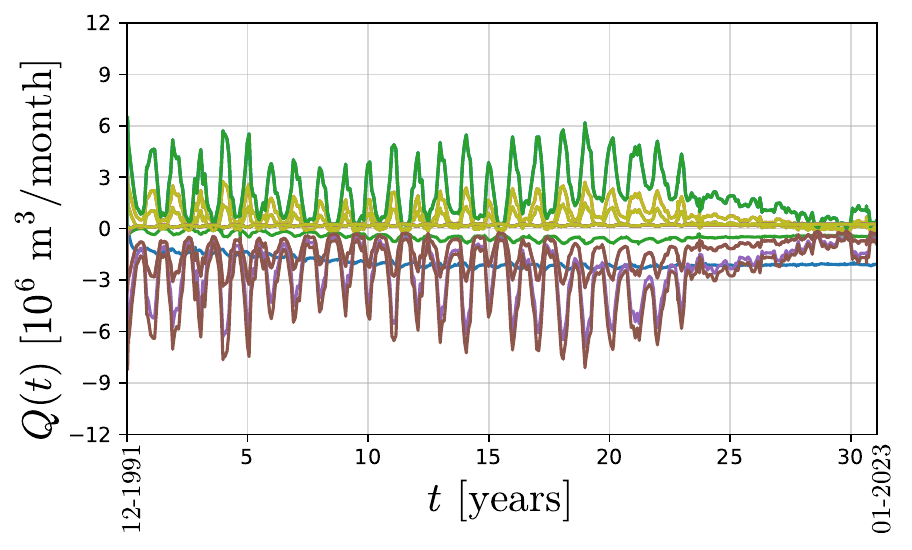}
  \includegraphics[width=6.5cm,keepaspectratio]{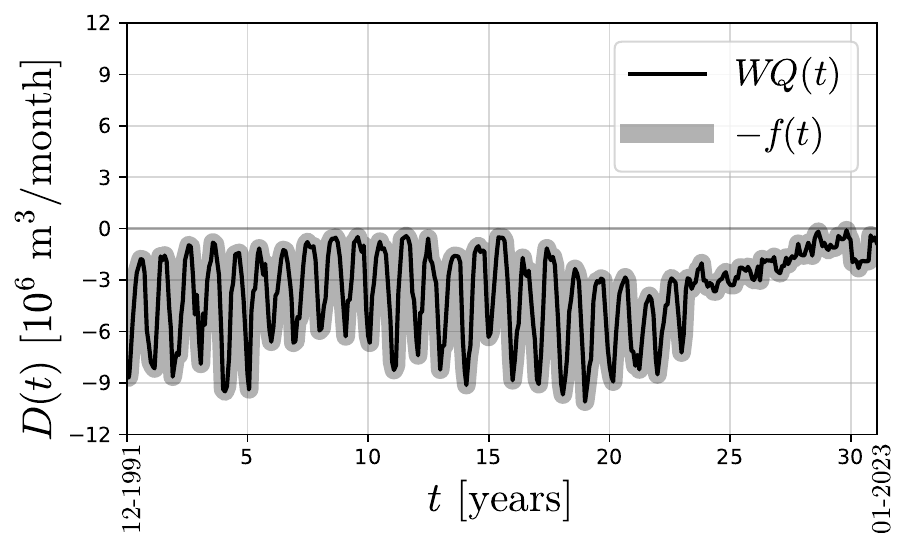}
  \caption{Control signals (top) and demand (bottom) in Scenario 1.}
  \label{fig:Q1}
\end{figure}

The results are displayed in Figs. \ref{fig:output1} and \ref{fig:Q1}. The control successfully drives all the pressure outputs to their smooth references. The average SR presents an overshoot in the beginning but then it remains close to the reference $R^*$. Therefore, after an initial transient, the control maintains the average SR close to $R^*$ in region $V_{R_1}$, in contrast with the historical uncontrolled scenario shown in Fig.~\ref{fig:validation}. Additionally, this was achieved while satisfying the gas extraction constraint, as set in \eqref{eq:restriction} and shown in Fig. \ref{fig:Q1}. The control signals remain within the order of magnitude of historical production rates at Groningen well clusters, and the observed oscillations are due to the demand signal rather than the control definition.

\subsection{Scenario 2: Gas extraction and CO$_2$ injection}

The same gains and parameters as in Scenario 1 were used in this case too, with the exception of the demand and weight matrix. We consider a hypothetical injection rate equal to approximately 1.36 times the gas extraction demand, from a simplified mass balance at 60 [MPa] and 100 [$^{\circ}$C]. This calculation does not establish carbon neutrality because it neglects fluid mixing, transport and the operational energy cost. Compensating injection was considered for Groningen in \citep{b:NAM2016GPM}, where nitrogen was proposed instead of CO$_2$. Here, CO$_2$ injection is used only as a conceptual example of separate injection and extraction channels. 

\begin{figure}[ht!]
  \centering
  \hspace{-8pt}\includegraphics[width=6.8cm,keepaspectratio]{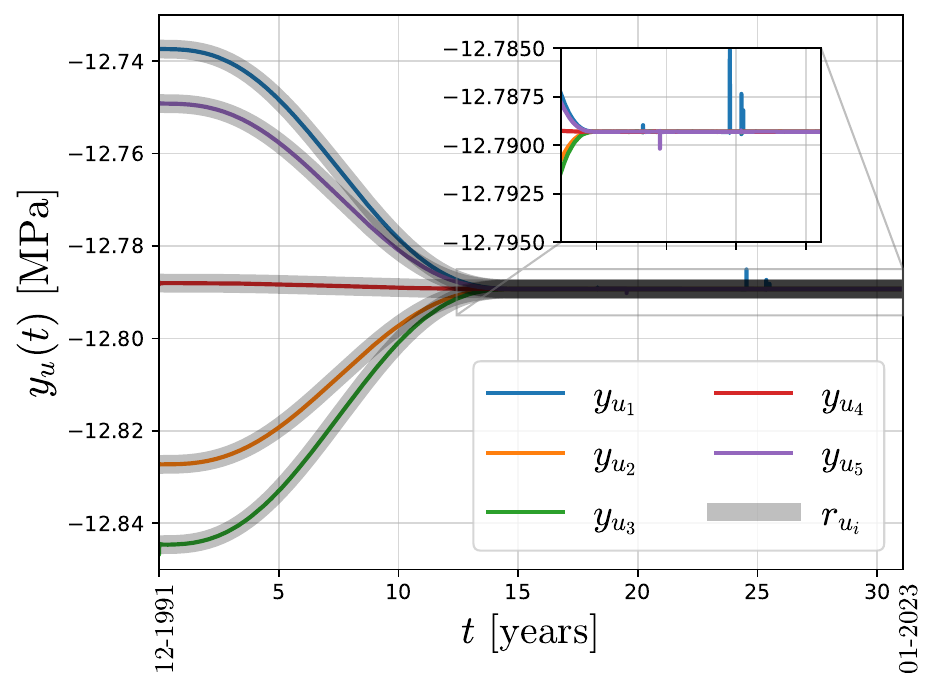}
  \includegraphics[width=6.5cm,keepaspectratio]{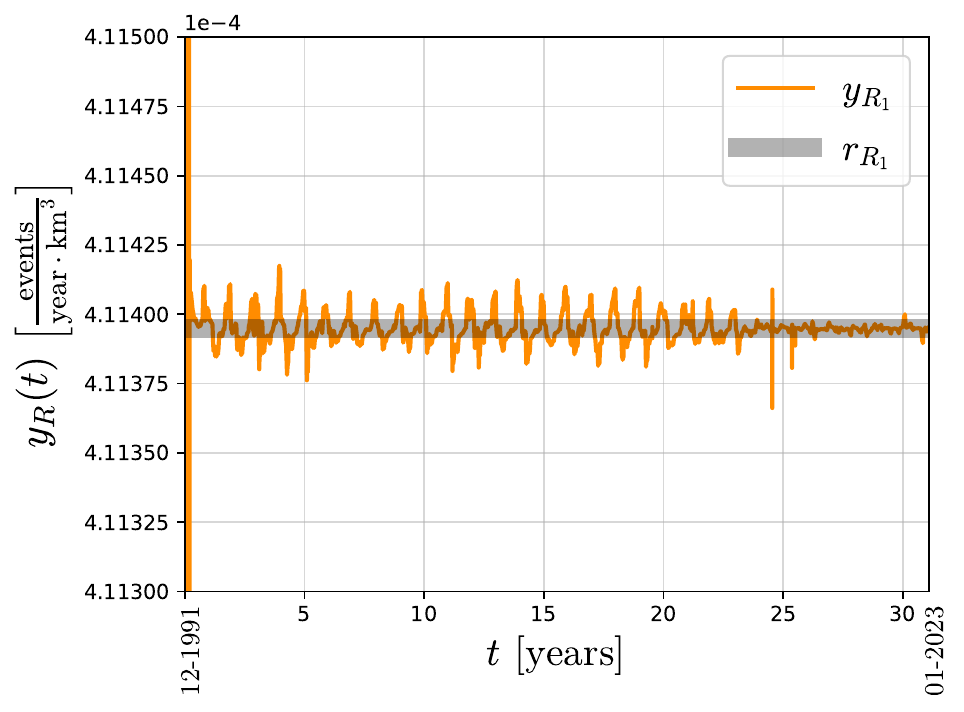}
  \caption{Pressure (top) and SR (bottom) outputs in Scenario 2.}
  \label{fig:output2}
\end{figure}

\begin{figure}[ht!]
  \centering
  \includegraphics[width=6.5cm,keepaspectratio]{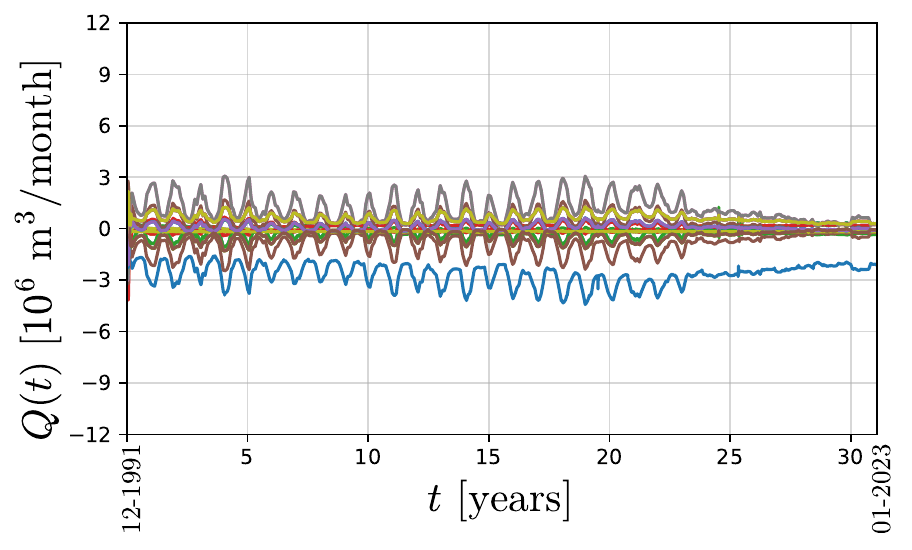}
  \includegraphics[width=6.5cm,keepaspectratio]{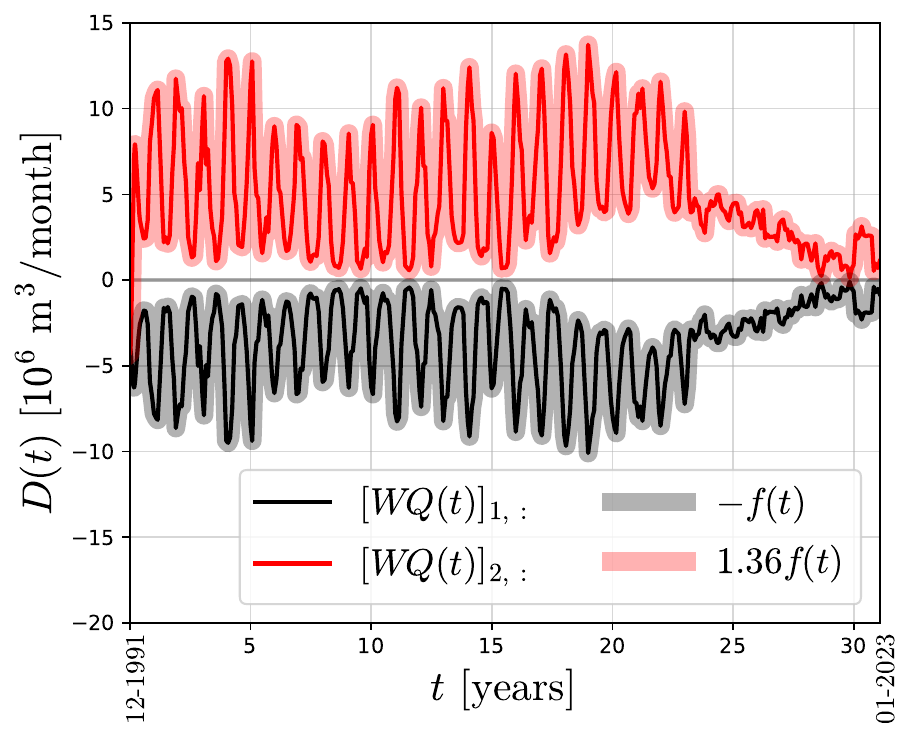}
  \caption{Control signals (top) and demand (bottom) in Scenario 2.}
  \label{fig:Q2}
\end{figure}

The demand is then defined by $D(t) = [-f(t),\,1.36f(t)]$, and the weight matrix $W \in \mathbb{R}^{2 \times 29}$ is chosen so that its first row distributes the demand among the first 14 components of $Q(t)$, whereas its second row distributes the demand among the last 15 components. Then, the same optimization process was performed to minimize $\|\overline{W}(B_0\overline{W})^{+}\|$ as before. The results are shown in Figs. \ref{fig:output2} and \ref{fig:Q2}. The control successfully drives both types of outputs (pressure and SR) to their respective references. The generated control signal exhibits lower saturation levels than in scenario 1, resulting in less control effort in this conceptual comparison. Lastly, both types of demands (injection and extraction) are followed at each time step.

\subsection{Comparison and limitations}

\begin{figure}[ht!]
  \centering
  \includegraphics[width=6.5cm,keepaspectratio]{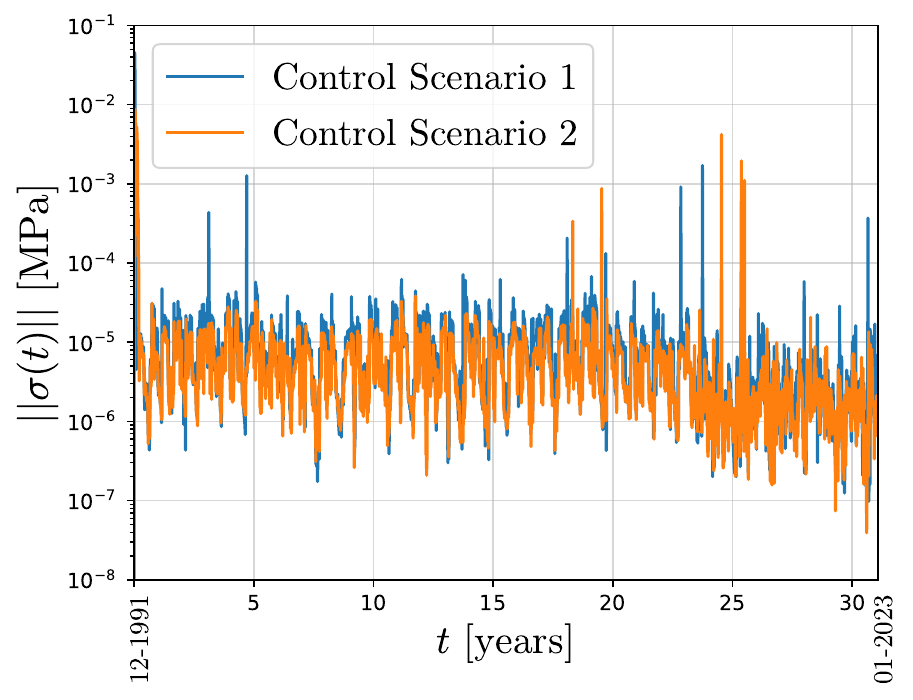}\\
  \includegraphics[width=6.3cm,keepaspectratio]{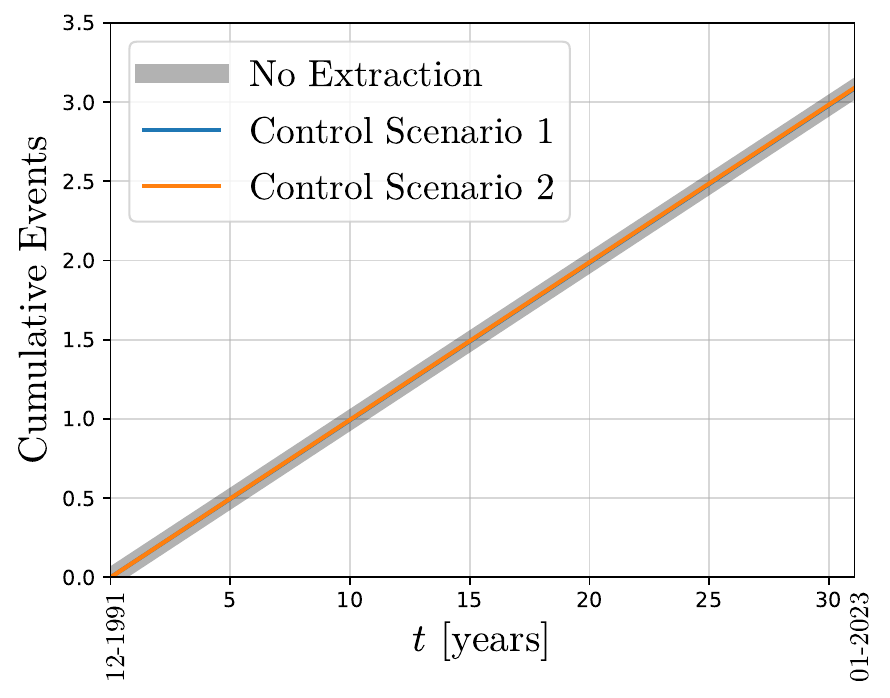}
  \caption{Error norm (top) and total cumulative events (bottom) in both scenarios.}
  \label{fig:enormT}
\end{figure}

The results of both scenarios will be compared. First, the Euclidean norm of the error is shown in Fig. \ref{fig:enormT} (top) (see Section \ref{sec:introduction} for the definition of the Euclidean norm). In both scenarios, the control strategy successfully stabilizes the error norm as expected, achieving the same level of precision. Second, Fig.~\ref{fig:enormT} (bottom) illustrates the cumulative number of events over the entire simulation period ($t_{max}\approx 31$~[years]), compared to the case without extraction, where the background SR predicts the cumulative events as $\int_V \int_0^{t_{max}} R^*(x) \, dt \, dV = R^* t_{max} |V|$, which is approximately $3$~[events]. Within this reduced model, both controlled scenarios approached the background cumulative count, compared with about 690 events for the historical no-control case (see Fig. \ref{fig:validation}). The results show
that the proposed control reduces the expected number of seismic events in the considered numerical model.

The model adopted in this work has several limitations. It neglects
detailed fault geometry, reservoir compaction, poroelastic stress
transfer and the full porodynamic coupling of \citep{b:Biot-1941}.
Furthermore, the theoretical model considers a known and spatially
homogeneous sign $s$, allowing injection to increase the SR for $s=1$
and extraction to increase it for $s=-1$. However, the stress response
to fluid operations depends on the local stress state, fault orientation,
reservoir compaction and poroelastic effects. Consequently, injection
may increase seismicity in some regions, while extraction may have the
opposite effect \citep{b:NAM2016GPM}. Spatially heterogeneous or unknown
and time-varying signs are not covered by the present control design. In \citep{b:Gutierrez-Stefanou-2026_c}, the problem of unknown and
time-varying control directions was addressed using a switching
sign-compensation mechanism for MIMO systems. This approach preserves
the nominal controller and recovers its stability properties once the
correct control direction is identified automatically by the control algorithm.

Another limitation concerns the use of the expected SR as the
control objective. Although regulating the SR can reduce the
expected number of seismic events, the controller does not
directly regulate earthquake magnitudes or provide bounds on
individual events. However, in a probabilistic sense, magnitude and seismicity rate are generally connected (see Gutenberg--Richter magnitude distribution and \citep{https://doi.org/10.1002/jgrb.50264,
b:https://doi.org/10.1029/2024GL110139}). For this purpose, in a related work
\citep{b:Gutierrez-Stefanou-2026_b}, a stochastic model was used for generating discrete in time and space seismic events with a probabilistically assigned magnitude.

Regarding the numerical validation, the pressure and SR models
were fitted to observations using a reduced two-dimensional
model. Such reduced-dimensional models have been applied to Groningen \citep{b:https://doi.org/10.1029/2020JB020013,doi:10.1144/SP528-2022-169,b:doi.org/10.1029/2023GL105455,BourneOates2017}. Their use is further motivated by the reservoir geometry, characterized by a relatively small thickness of 0.1--0.3 km compared with its lateral extent, covering approximately 900 km$^2$. Although the resulting model reproduces the main
seismicity trends observed in Groningen, it does not constitute
predictive validation of the full three-dimensional reservoir.
Related but simpler controllers were tested under strong heterogeneity of
hydraulic diffusivity and compressibility in academic examples
\citep{b:Gutierrez-Stefanou-2024}. Numerical sensitivity studies
of the control gains, sampling and saturation were also performed
in \citep{b:Gutierrez-Stefanou-2026_b}. 

The theoretical analysis also considers continuous-time measurements and actuation. In practical applications, measurement delays, discrete control updates and well-rate limits may affect the control performance. Some of these effects were investigated numerically in \citep{b:Gutierrez-Stefanou-2026_b}, where stochasticity, filtered event-count feedback, zero-order-hold updates, saturation and gain selection were considered. Moreover, the weighted demand constraint $WQ=D$ does not necessarily impose a prescribed physical net extraction, as discussed in Section~4.2. Ongoing work aims to address this through a constrained optimization problem that balances SR regulation, well-rate saturation limits, and the imposition of a prescribed net extraction profile.

Finally, Scenario~2 considers CO$_2$ injection only as a
conceptual example of the control allocation. The model does
not account for multiphase flow, gas mixing, CO$_2$--rock
reactions, changes in hydraulic properties, injectivity or
well integrity. Therefore, this scenario does not establish
the feasibility of CO$_2$ injection at Groningen or demonstrate
carbon neutrality. Future work should address these limitations,
including more detailed fault-scale stresses, magnitude-dependent
control objectives and measurement delays in the closed-loop
analysis.

\section{Conclusions}
\label{sec:conclusion}

This paper introduces a robust control strategy for output tracking of a nonlinear PDE-ODE system in $n_d\in\{1,2,3\}$ spatial dimensions where the ODE has a logistic-like dynamics. The output feedback control was developed through a mathematical analysis of the cascade system. First, bounds on the solutions and their time derivatives were established for the diffusion equation and the nonlinear ODE system. These bounds were then used to prove the boundedness of the uncertain control coefficient and the time derivative of the perturbation. The mathematical formulation demonstrates the controller's capability to achieve tracking for two types of outputs in the system, even in the presence of heterogeneities in the system, model uncertainties, and under limited system information, all while using a continuous control signal. This study is motivated by the challenging problem of mitigating human-induced seismicity in geoenergy applications, but we believe that the proposed methodology can also be applied to various challenging applications using similar PDE-ODE dynamics.

A case study inspired by the Groningen gas reservoir illustrates the proposed control under production-demand constraints. The numerical simulations use a reduced 2D model fitted to Groningen observations. Under the stated assumptions, the control approach ensures tracking of desired regional SR and pressures despite admissible uncertainties. This is particularly valuable in such complex systems, where real parameters (such as diffusivity and compressibility) are often difficult to acquire with high precision. Moreover, the control design effectively addressed the complex interconnection between the infinite-dimensional system and the nonlinear ODE for $n_d\in\{1,2,3\}$.

Numerical simulations illustrate a reduction in the expected SR of the reduced model and compare two idealized well-allocation scenarios. While our theoretical approach has limitations, robust control theory offers new possibilities for addressing the challenges of uncertain, nonlinear distributed parameter systems, providing a framework for balancing seismicity mitigation with renewable energy production and storage objectives.

Ongoing work \citep{b:Gutierrez-Stefanou-2026_b,b:Gutierrez-Stefanou-2026_c} focuses on handling unknown and time-varying control directions, as well as integrating SR regulation, well-rate saturation limits and prescribed net extraction within a constrained optimization framework. Future developments should address spatially heterogeneous SR responses, measurement delays and sampled-data stability, together with more realistic three-dimensional reservoir models incorporating poroelastic effects and fault geometry. Magnitude-dependent control objectives and multiphase flow modeling would further extend the approach toward practical field applications.

\backmatter

%
%
%

\bmhead{Acknowledgements}

Funded by the European Union. Views and opinions expressed are however those of the author(s) only and do not necessarily reflect those of the European Union or the European Research Council Executive Agency.  Neither the European Union nor the granting authority can be held responsible for them. This work is supported by the ERC grant INJECT, no. 101087771, doi: 10.3030/101087771. The first author would also like to thank the Region Pays de la Loire and Nantes M\'etropole for their support under the Connect Talent programme (CEEV: Controlling Extreme EVents - Blast: Blas LoAds on STructures). Both authors would like to express their sincere thanks to Dr. Mateo Acosta, Prof. J. P. Avouac, Dr. Stephen Bourne and Dr. Jan van Elk for their fruitful discussions about the Groningen reservoir and related data.

\section*{Declarations}

\subsection*{Funding}
Funded by the European Union. Views and opinions expressed are however those of the author(s) only and do not necessarily reflect those of the European Union or the European Research Council Executive Agency.  Neither the European Union nor the granting authority can be held responsible for them. This work is supported by the ERC grant INJECT, no. 101087771, doi: 10.3030/101087771.

\subsection*{Disclaimer}

This study is conducted for academic purposes only. The approaches described are not intended for direct field application or operational decision-making without site-specific validation, calibration, and expert review under applicable safety and regulatory frameworks. Any use beyond this academic context remains the responsibility of the implementing parties.

\subsection*{Declaration of generative AI and AI-assisted technologies}

During the preparation of this work, the authors used ChatGPT for language editing and to assist in the construction of the Lean~4 formalization. The authors reviewed and edited the generated material before inclusion in the manuscript. 

\subsection*{Competing interests}
None declared.

\subsection*{Author contributions}
\textbf{Diego Guti\'errez-Oribio:} Conceptualization, Formal Analysis, Methodology, Software, Validation, Visualization, Writing - Original Draft Preparation. \textbf{Ioannis Stefanou:} Conceptualization, Formal Analysis, Funding Acquisition, Methodology, Project Administration, Software, Writing - Review \& Editing

\subsection*{Data availability}
The data are available from the corresponding author on reasonable request.

\begin{appendices}

\section{Proof of Theorem \ref{th:1}}
\label{app:control}

We prove the bounds in \eqref{eq:DeltaBbound}--\eqref{eq:Psi2bound}
needed to apply the MIMO super-twisting result of
\citep{b:Mathey-Moreno-2024}. We consider a forward-complete
solution satisfying Assumptions~\ref{A1}--\ref{A5}.
First, we use an energy estimate and semigroup theory to
establish bounds on the pressure and its time derivative.
Then, we transform the SR equation into a logistic equation
and use comparison arguments to bound the SR and its time
derivative, together with its spatial regularity near the
wells. Next, we bound the time derivative of the regional perturbation, $\Psi(t)$, in
the tracking-error dynamics by combining spatial regularity
and semigroup estimates. Finally, we use these results to bound the
uncertain control coefficient $G(t)$ and the time derivative
of the normalized perturbation $\bar\Psi(t)$.

The notation $D(A)$ denotes the domain of the diffusion operator in Assumption~\ref{A5}. All bounds below are uniform for $t\geq0$ unless stated otherwise. We use $C$ for positive constants that may change from one estimate to another but do not depend on time. The required semigroup, Sobolev, and elliptic estimates can be found in \citep{b:EvansPDE,b:Pazy,b:Ouhabaz2005}.

We will use H\"older's inequality $\int_U |fg|\,dV\leq\|f\|_{L^p(U)}\|g\|_{L^q(U)}$, where $1/p+1/q=1$. In particular, for $p=q=2$,
we obtain the Cauchy--Schwarz inequality $|\langle f,g\rangle|\leq\|f\|_2\|g\|_2$. We also use Young's inequality $2ab\leq\varepsilon a^2+\varepsilon^{-1}b^2$ for $\varepsilon>0$. For a mean-zero function $v\in H^1(V)$, the Poincar\'e--Wirtinger inequality gives $\|v\|_2\leq C_P\|\nabla v\|_2$. On bounded Lipschitz subsets of $\mathbb R^{n_d}$, $n_d\in\{1,2,3\}$, we also use $H^2\hookrightarrow W^{1,6} \hookrightarrow L^\infty$ and the fact that products of $H^2$ functions remain in $H^2$.

\subsection{Boundedness of the diffusion process}
\label{app:control_bound_diff}

Since the Neumann diffusion operator does not dissipate constant functions, we separate the mean pressure from the mean-zero part. Set $p(x,t)=u(x,t)-\bar u(t)$, where $\bar u(t)=|V|^{-1}\int_Vu\,dV$. Integrating \eqref{eq:diff} over $V$ and using the Neumann boundary condition yields
\begin{equation}
 \dot{\bar u}(t)=\frac{1}{\beta|V|}\sum_{j=1}^nQ_j(t),\qquad
 p_t=-Ap+\sum_{j=1}^nf_jQ_j(t),
 \label{eq:diff2}
\end{equation}
where $f_j=\beta^{-1}(\mathcal B_j-|V|^{-1})\in L^2(V)$ has zero spatial average. The operator $A$ is defined in Assumption~\ref{A5}. It is nonnegative and self-adjoint in $L^2(V)$, and its kernel consists of constant functions. On the mean-zero subspace, $A$ has a positive spectral gap $\omega>0$. Its semigroup is written as $e^{-At}$. The domain $D(A)$ consists of $H^1(V)$ functions with an $L^2(V)$ diffusion operator and the homogeneous conormal boundary condition in the weak sense. 

We first establish an energy estimate. Taking the $L^2$ inner product of the second equation in \eqref{eq:diff2} with $p$ gives
\begin{equation*}
 \frac12\frac{d}{dt}\|p\|_2^2
 =-\langle p,Ap\rangle+\sum_{j=1}^n\langle p,f_j\rangle Q_j.
\end{equation*}
By Assumption~\ref{A3}, $\langle p,Ap\rangle\geq k^m(\beta\eta^M)^{-1}\|\nabla p\|_2^2$. Since $p$ has zero average, the Poincar\'e--Wirtinger inequality gives $\langle p,Ap\rangle\geq\omega\|p\|_2^2$. Writing $F=(\sum_j\|f_j\|_2^2)^{1/2}$, Cauchy--Schwarz and Young's inequality give
\begin{equation*}
 \frac{d}{dt}\|p\|_2^2\leq-\omega\|p\|_2^2+\frac{F^2}{\omega}\|Q(t)\|^2.
 \label{eq:pEnergyNew}
\end{equation*}
Integration and Assumption~\ref{A1} yield
\begin{equation}
 \|p(t)\|_2^2\leq e^{-\omega t}\|p(0)\|_2^2
 +\frac{F^2L_Q^2}{\omega^2}(1-e^{-\omega t}).
 \label{eq:pISSNew}
\end{equation}
This is an exponential input-to-state bound for the mean-zero pressure.

Next, we bound $Ap$, which is needed to bound $p_t$ and the regional terms in \eqref{eq:Psi}. The mild solution of \eqref{eq:diff2} is $p(t)=e^{-At}p(0)+\sum_j\int_0^te^{-A(t-s)}f_jQ_j(s)\,ds$. Let $\lambda_k>0$ denote the positive eigenvalues of $A$ and let $(f_j)_k$ be the corresponding coefficients of $f_j$. For every $j$ and $k$,
\begin{equation*}
 \left|\int_0^t\lambda_ke^{-\lambda_k(t-s)}(f_j)_kQ_j(s)\,ds\right|
 \leq |(f_j)_k|\|Q_j\|_{L^\infty(0,\infty)},
\end{equation*}
since $\int_0^t\lambda_ke^{-\lambda_k(t-s)}\,ds\leq1$. Summing the squares over the eigenfunctions and then adding the contributions of all wells gives
\begin{equation}
 \sup_{t\geq0}\|Ap(t)\|_2
 \leq\|Ap(0)\|_2+\sum_{j=1}^n\|f_j\|_2\|Q_j\|_{L^\infty(0,\infty)}<\infty.
 \label{eq:ApboundNew}
\end{equation}
Here $p(0)\in D(A)$ by Assumption~\ref{A5}. This estimate uses the fixed spatial coefficients $f_j$ and only the bounds on $Q_j$, without differentiating the well rates. The second equation in \eqref{eq:diff2} now shows that $\sup_t\|p_t(t)\|_2<\infty$. Together with \eqref{eq:pISSNew}, this proves that $p$ is uniformly bounded in $D(A)$.

For $n_d\in\{1,2,3\}$, the Neumann heat-semigroup estimate $\|e^{-At}\|_{L^2\to L^\infty}\leq C(1+t^{-n_d/4})$ implies $D(A)\hookrightarrow L^\infty(V)$. Indeed, writing $(I+A)^{-1}=\int_0^\infty e^{-s}e^{-As}\,ds$, the factor $s^{-n_d/4}$ is integrable at zero for $n_d<4$. Hence $\sup_t\|p(t)\|_\infty<\infty$. Interior elliptic regularity, using the local $W^{1,\infty}$ regularity in Assumption~\ref{A3}, also gives $\sup_t\|p(t)\|_{H^2(U_j)}<\infty$.

The absolute pressure is bounded directly by the accumulated-flux condition in Assumption~\ref{A1}. Integrating the first equation in \eqref{eq:diff2} gives $\bar u(t)=\bar u(0)+(\beta|V|)^{-1}\int_0^t\sum_jQ_j(s)\,ds$, and therefore $\sup_t|\bar u(t)|\leq |\bar u(0)|+L_{\mathrm{net}}/(\beta|V|)$. Since $u=p+\bar u$ and $u_t=p_t+\dot{\bar u}$, we conclude that
\begin{equation}
 \sup_{t\geq0}\left(\|u(t)\|_{L^\infty(V)}+\|u_t(t)\|_{L^2(V)}+\|u(t)\|_{H^2(U_j)}\right)<\infty.
 \label{eq:uUniformNew}
\end{equation}

For homogeneous Dirichlet boundary conditions, $u=0$ on $\partial V$, the corresponding diffusion operator has no
constant mode and has a positive spectral gap on $L^2(V)$. Therefore, the energy and semigroup estimates above can be
applied directly to $u$, without separating its spatial mean. Under the corresponding Dirichlet regularity and initial
compatibility conditions, these estimates recover the bound \eqref{eq:uUniformNew}. Consequently, the remaining arguments in this proof apply in the same way.

\subsection{Boundedness of the SR dynamics}
\label{app:control_bound_SR}

To eliminate $u_t$ from the SR equation, define $S=Re^{-s\gamma_1u}$. Differentiating and using \eqref{eq:SR} gives 
\begin{equation}
 S_t=\gamma_2R^*S-\gamma_2e^{s\gamma_1u}S^2.
 \label{eq:SRtransNew}
\end{equation}
The boundedness of $u$ in \eqref{eq:uUniformNew} and Assumption~\ref{A3} give positive constants $a_m,a_M,b_m,b_M$ such that $a_m\leq\gamma_2R^*\leq a_M$ and $b_m\leq\gamma_2e^{s\gamma_1u}\leq b_M$ on $V\times T$.

We now give the comparison estimates. The initial value $S(0)=R^0e^{-s\gamma_1u^0}$ is bounded above on $V$ by Assumption~\ref{A5} and \eqref{eq:uUniformNew}. The differential inequality $S_t\leq a_MS-b_mS^2$ yields
\begin{equation*}
 0<S(x,t)\leq \max\left\{\|S(0)\|_{L^\infty(V)},\frac{a_M}{b_m}\right\}
 \quad\text{for almost every }x\in V.
\end{equation*}
For each $U_j$, Assumption~\ref{A5} also gives $S(x,0)\geq s_j^m>0$. Using $S_t\geq a_mS-b_MS^2$ on $U_j$ gives
\begin{equation*}
 S(x,t)\geq\min\left\{s_j^m,\frac{a_m}{b_M}\right\}>0
 \quad\text{for almost every }x\in U_j.
\end{equation*}
Since $R=Se^{s\gamma_1u}$ and $\gamma_1u$ is bounded, these inequalities prove $\sup_t\|R(t)\|_\infty<\infty$ and a positive lower bound for $R$ on each $U_j$.

We next establish the spatial bound needed near the wells. The change $Y=S^{-1}$ converts \eqref{eq:SRtransNew} into a linear equation, whose spatial derivatives can be bounded directly. Write $a=\gamma_2R^*$ and $b=\gamma_2e^{s\gamma_1u}$. Equation \eqref{eq:SRtransNew} becomes
\begin{equation}
 Y_t=-aY+b.
 \label{eq:SRrecNew}
\end{equation}
On each $U_j$, the local assumptions on $\gamma_1$, $\gamma_2$, and $R^*$, together with \eqref{eq:uUniformNew}, give uniform $H^2(U_j)$ bounds for $a$ and $b$. This follows from the $H^2$ product property for $n_d\in\{1,2,3\}$. We already know that $a\geq a_m>0$ and $Y$ is uniformly bounded above and away from zero. Also, $Y(0)=e^{s\gamma_1u^0}/R^0\in H^2(U_j)$ because $R^0$ is positive and belongs to $H^2(U_j)$ by Assumption~\ref{A5}.

Differentiating \eqref{eq:SRrecNew} once in space gives $(\nabla Y)_t=-a\nabla Y-Y\nabla a+\nabla b$. The variation-of-constants formula, applied pointwise in space, yields
\begin{equation*}
 \nabla Y(t)=e^{-\int_0^ta(s)\,ds}\nabla Y(0)
 +\int_0^t e^{-\int_s^ta(\tau)\,d\tau}\bigl(\nabla b(s)-Y(s)\nabla a(s)\bigr)\,ds.
\end{equation*}
The exponentials are bounded by $e^{-a_m(t-s)}$. Since $H^2(U_j)\hookrightarrow W^{1,6}(U_j)$, both $\nabla a$ and $\nabla b$ are uniformly bounded in $L^6(U_j)$. The bound on $Y$ then gives $\sup_{t\geq0}\|\nabla Y(t)\|_{L^6(U_j)}<\infty$. 

Differentiating once more gives $(D^2Y)_t=-aD^2Y-\nabla a\otimes\nabla Y
 -\nabla Y\otimes\nabla a-YD^2a+D^2b$. Each forcing term is uniformly bounded in $L^2(U_j)$. For example, since $U_j$ is bounded, $\|\nabla a\otimes\nabla Y\|_{L^2(U_j)}\leq |U_j|^{1/6}\|\nabla a\|_{L^6(U_j)}\|\nabla Y\|_{L^6(U_j)}$. The same variation-of-constants estimate, now in $L^2(U_j)$, gives $\sup_t\|D^2Y(t)\|_{L^2(U_j)}<\infty$. Thus $Y$ is uniformly bounded in $H^2(U_j)$. Since $Y$ is bounded away from zero and $R=e^{s\gamma_1u}/Y$, the $H^2$ product property gives $\sup_{t\geq0}\|R(t)\|_{H^2(U_j)}<\infty$. Finally, using \eqref{eq:SR} and the bounds already obtained for $R$, $u_t$, $\gamma_1$, $\gamma_2$, and $R^*$, imply
\begin{equation}
\sup_{t\geq0}\left(
\norm{R(t)}_\infty+
\norm{R_t(t)}_2+
\norm{R(t)}_{H^2(U_j)}
\right)<\infty.
\label{eq:RUniformNew}
\end{equation}

\subsection{Regional perturbation derivatives}
\label{app:control_bound_pert}

We now bound $\dot\Psi$ without assuming that $\dot Q$ is bounded. The main difficulty comes from the terms containing $Ap$ in \eqref{eq:Psi}. Let $\chi_E$ denote the characteristic function of a measurable set $E$, equal to one on $E$ and zero elsewhere. For a pressure region, set $w=\chi_{V_{u_i}}/|V_{u_i}|$. For an SR region, set $w=\chi_{V_{R_i}}\gamma_1R/(\gamma_{1_0}R_0^*|V_{R_i}|)$. Then the diffusion terms in \eqref{eq:Psi} are $-\langle w,Ap\rangle$ for both pressure and SR. Equations \eqref{eq:ApboundNew} and \eqref{eq:RUniformNew} give $\sup_t(\|w(t)\|_2+\|w_t(t)\|_2)<\infty$. For the pressure weights, $w_t=0$. For the SR weights, $w_t=\chi_{V_{R_i}}\gamma_1R_t/(\gamma_{1_0}R_0^*|V_{R_i}|)$, which is bounded by the $L^2$ estimate on $R_t$.

A direct differentiation of $\langle w,Ap\rangle$ would involve $Ap_t$ and, through the pressure equation, $\dot Q$. Moreover, the characteristic functions jump at the output-region boundaries, so the full weights need not belong to $D(A)$. We therefore divide each observation into a part near the wells and a part away from them. Assumption~\ref{A4} gives a positive distance between each well source region and all output-region boundaries. Choose a smooth function $\zeta$ equal to one near every well source region and supported away from these boundaries and from $\partial V$. On the part where an SR weight is nonzero, the support of $\zeta$ is contained in the neighborhoods $U_j$ from Assumption~\ref{A3}. We write $w=\zeta w+(1-\zeta)w$.

On the support of $\zeta$, the pressure weight is either constant or zero. The SR weight has the local $H^2$ regularity established in \eqref{eq:RUniformNew}. The local $W^{1,\infty}$ regularity of $k/\eta$ and the fact that $\zeta w$ vanishes near $\partial V$ then imply $\zeta w\in D(A)$, with a uniform bound on $\|A(\zeta w)\|_2$. Consequently, for $J_{\mathrm{near}}=\langle\zeta w,Ap\rangle$ we have, in the weak sense,
\begin{equation*}
 \dot J_{\mathrm{near}}=\langle(\zeta w)_t,Ap\rangle+\langle A(\zeta w),p_t\rangle.
\end{equation*}
Cauchy--Schwarz, \eqref{eq:ApboundNew}, and the bound on $p_t$ give $\sup_t|\dot J_{\mathrm{near}}(t)|<\infty$. Thus the near part does not require $\dot Q$.

For the far part, $(1-\zeta)w$ is at a fixed positive distance $d_{\mathrm{sep}}>0$ from every well source region. This separation allows us to estimate the effect of each well using the heat semigroup. Fix the observation time $t$ and define the following kernel for $\tau>0$, keeping $w(t)$ fixed,
\begin{equation*}
 K_j(\tau)=\left\langle(1-\zeta)w(t),Ae^{-A\tau}\mathcal B_j/\beta\right\rangle.
\end{equation*}
Because the two supports are separated, off-diagonal heat-semigroup estimates give $|K_j(\tau)|\leq C\tau^{-1}e^{-cd_{\mathrm{sep}}^2/\tau}$ and $|K_j'(\tau)|\leq C\tau^{-2}e^{-cd_{\mathrm{sep}}^2/\tau}$ for $0<\tau\leq1$. The prime denotes differentiation in $\tau$. For $\tau\geq1$, the spectral gap gives exponential decay of both kernels, since $A$ annihilates constants. The constants are uniform in $t$ because $\sup_t\|w(t)\|_2<\infty$. In particular, the kernel vanishes as $\tau\to0$ and its derivative is integrable. Thus
\begin{equation}
 K_j(0)=0,\qquad
 \sup_{t\geq0}\left(\sup_{\tau>0}|K_j(\tau)|+
 \int_0^\infty|K_j'(\tau)|\,d\tau\right)<\infty.
 \label{eq:farKernelNew}
\end{equation}
These estimates follow from the Davies--Gaffney bounds for
uniformly elliptic diffusion operators and the analyticity
of their semigroups \citep[Section 6.3 and 6.5]{b:Ouhabaz2005}. 

Differentiation of \eqref{eq:diff2} gives $p_t(t)=e^{-At}p_t(0)+\sum_j\int_0^te^{-A(t-s)}f_j\dot Q_j(s)\,ds$. The condition $p_t(0)\in D(A)$ ensures that $Ae^{-At}p_t(0)$ is bounded. The constant part of $f_j$ does not contribute because $A$ annihilates constants. By integration by parts, $\int_0^tK_j(t-s)\dot Q_j(s)\,ds=-K_j(t)Q_j(0)+\int_0^tK_j'(t-s)Q_j(s)\,ds$, where we used $K_j(0)=0$. Consequently, the time derivative of
$J_{\mathrm{far}}(t)=\langle(1-\zeta)w(t),Ap(t)\rangle$
is given by
\begin{equation}
\begin{split}
\dot J_{\mathrm{far}}(t)
={}&\left\langle(1-\zeta)w_t(t),Ap(t)\right\rangle\\
&+\left\langle(1-\zeta)w(t),Ae^{-At}p_t(0)\right\rangle\\
&-\sum_{j=1}^n K_j(t)Q_j(0)\\
&+\sum_{j=1}^n\int_0^t K_j'(t-s)Q_j(s)\,ds.
\end{split}
\label{eq:farDerivativeNew}
\end{equation}
The first term is bounded by \eqref{eq:uUniformNew} and the bound on \(w_t\). The second term is bounded by $\|(1-\zeta)w(t)\|_2\|Ap_t(0)\|_2$, which is finite by Assumption~\ref{A5}. The other terms are bounded by \eqref{eq:farKernelNew} and Assumption~\ref{A1}. The variation of the observation weight adds $\langle[(1-\zeta)w]_t,Ap\rangle$, which is bounded by \eqref{eq:ApboundNew}. Combining the near and far parts shows that $J(t)=\langle w(t),Ap(t)\rangle$ belongs to $W^{1,\infty}(0,\infty)$.

We now return to \eqref{eq:Psi}. Let $J_{u_i}$ and $J_{R_i}$ denote $J=\langle w,Ap\rangle$ for the pressure and SR weights introduced above, respectively. The bounds on $w$ and $Ap$ imply that $J_{u_i}$ and $J_{R_i}$ are uniformly bounded. Moreover, the near and far estimates show that their time derivatives are also uniformly bounded. Hence, the components of \eqref{eq:Psi} can be written as
\begin{equation*}
\begin{split}
 [\psi_1(t)]_i&=-J_{u_i}(t)-\dot r_{u_i}(t),\\
 [\psi_2(t)]_i&=-J_{R_i}(t)-\frac{s}{\gamma_{1_0}R_0^*|V_{R_i}|}
 \int_{V_{R_i}}\gamma_2 R(R-R^*)\,dV\\
 &\quad-\frac{s\dot r_{R_i}(t)}{\gamma_{1_0}R_0^*}.
\end{split}
\end{equation*}
By Assumption~\ref{A3} and \eqref{eq:RUniformNew}, the integral in the SR component is uniformly bounded. Assumption~\ref{A2} bounds the first derivatives of the references. Consequently, $\Psi(t)$ is uniformly bounded.

To bound its time derivative, we differentiate the preceding expressions. Since $R^*$ does not depend on time, we obtain
\begin{equation*}
\begin{split}
 [\dot\psi_1(t)]_i&=-\dot J_{u_i}(t)-\ddot r_{u_i}(t),\\
 [\dot\psi_2(t)]_i&=-\dot J_{R_i}(t)-\frac{s}{\gamma_{1_0}R_0^*|V_{R_i}|}
 \int_{V_{R_i}}\left[\gamma_{2,t}R(R-R^*)
 +\gamma_2(2R-R^*)R_t\right]dV\\
 &\quad-\frac{s\ddot r_{R_i}(t)}{\gamma_{1_0}R_0^*}.
\end{split}
\end{equation*}
The derivatives of $J_{u_i}$ and $J_{R_i}$ are bounded as shown above. For the remaining integral, Assumption~\ref{A3} and H\"older's inequality give 
\begin{equation*}
\begin{split}
 &\left|\int_{V_{R_i}}\left[\gamma_{2,t}R(R-R^*)
 +\gamma_2(2R-R^*)R_t\right]dV\right|\\
 &\quad\leq L_{\gamma_{2,t}}|V_{R_i}|\|R(t)\|_\infty
 (\|R(t)\|_\infty+R_M^*)\\
 &\qquad+\gamma_2^M|V_{R_i}|^{1/2}
 (2\|R(t)\|_\infty+R_M^*)\|R_t(t)\|_2.
\end{split}
\end{equation*}
The right-hand side of this inequality is uniformly bounded by \eqref{eq:RUniformNew}. Assumption~\ref{A2} also bounds the second derivatives of the references. Therefore, each component of $\dot\Psi$ is bounded. Since there are finitely many output regions, we conclude that
\begin{equation}
 \sup_{t\geq0}\bigl(\|\Psi(t)\|+\|\dot\Psi(t)\|\bigr)<\infty.
 \label{eq:PsiDerivativeNew}
\end{equation}

\subsection{Boundedness of the uncertain coefficient and normalized perturbation}
\label{app:control_bound_coeff}

We first show the structure of $G=BB_0^+$. By Assumption~\ref{A4}, the nonzero entries of each row of $B_0$ in \eqref{eq:B01} correspond to wells in one output region, and distinct rows have disjoint sets of well indices. Therefore, $B_0B_0^T$ is diagonal and invertible.The matrix $B$ in \eqref{eq:B1} has nonzero entries at the same positions as $B_0$. Thus $BB_0^T$ is also diagonal. Since $B_0^+=B_0^T(B_0B_0^T)^{-1}$, the matrix $G=BB_0^+$ is diagonal.

For a pressure row, \eqref{eq:B1}--\eqref{eq:B01} give the diagonal entry $\beta_0/\beta$. For the $i$th SR row, let $\mathcal I_i=\{j:V_j^*\subset V_{R_i}\}$ denote the set of well indices in $V_{R_i}$, and let $N_i=|\mathcal I_i|$ be their number. The quantity $H_i(t)$ measures the mean SR sensitivity to the wells inside region $V_{R_i}$. Define it by
\begin{equation}
 H_i(t)=\frac1{N_i}\sum_{j\in \mathcal I_i}\frac1{|V_j^*|}\int_{V_j^*}\gamma_1(x)R(x,t)\,dV.
 \label{eq:SRavg}
\end{equation}
The nominal SR row has the same nonzero value $1/(\beta_0|V_{R_i}|)$ for each such well. Multiplying the actual row by its nominal right-pseudoinverse column gives $\beta_0H_i(t)/(\beta\gamma_{1_0}R_0^*)$. Hence
\begin{equation*}
 G(t)=I_m+\Delta B(t)
 =\operatorname{diag}\!\left(\frac{\beta_0}{\beta}I_{m_u},
 \left(\frac{\beta_0H_i(t)}{\beta\gamma_{1_0}R_0^*}\right)_{i=1}^{m_R}\right).
 \label{eq:DeltaB2}
\end{equation*}

The positive lower and upper bounds for $R$ on $U_j$, together with Assumptions~\ref{A3}--\ref{A5}, imply $0<H_m\leq H_i(t)\leq H_M<\infty$ for constants independent of $i$ and $t$. Let $0<\beta_0<2\beta$. Taking $\gamma_{1_0}R_0^*=(\beta_0/\beta)(H_m+H_M)/2$ gives
\begin{equation}
 \|G(t)-I_m\|\leq
 \max\!\left\{\left|\frac{\beta_0}{\beta}-1\right|,
 \frac{H_M-H_m}{H_M+H_m}\right\}
 =:\delta_B<1.
 \label{eq:DeltaBnorm}
\end{equation}
More generally, the condition
\begin{equation}
 0<\beta_0<2\beta,\qquad
 \gamma_{1_0}R_0^*>\frac{\beta_0}{2\beta}H_M
 \label{eq:constants}
\end{equation}
ensures $\|G-I_m\|<1$ whenever $H_m>0$. 

It remains to bound the derivative of $\bar\Psi=G^{-1}\Psi$. Differentiating \eqref{eq:SRavg} gives
\begin{equation*}
 \dot H_i(t)=\frac1{N_i}\sum_{j\in \mathcal I_i}
 \frac1{|V_j^*|}\int_{V_j^*}\gamma_1(x)R_t(x,t)\,dV.
\end{equation*}
Cauchy--Schwarz yields $|\dot H_i(t)|\leq (\gamma_1^M/N_i)\sum_{j\in \mathcal I_i}|V_j^*|^{-1/2}\|R_t(t)\|_{L^2(V_j^*)}$. The right-hand side is uniformly bounded by \eqref{eq:RUniformNew}. Thus $\sup_t\|\dot G(t)\|<\infty$. By \eqref{eq:DeltaBnorm}, all singular values of $G(t)$ are at least $1-\delta_B$, so $\|G(t)^{-1}\|\leq(1-\delta_B)^{-1}$.

Differentiating $G^{-1}G=I_m$ gives $\frac{d}{dt}G^{-1}=-G^{-1}\dot G G^{-1}$. Therefore,
\begin{equation}
 \dot{\bar\Psi}=G^{-1}\dot\Psi-G^{-1}\dot G G^{-1}\Psi,
 \qquad
 \|\dot{\bar\Psi}\|\leq
 \frac{\|\dot\Psi\|}{1-\delta_B}
 +\frac{\|\dot G\|\,\|\Psi\|}{(1-\delta_B)^2}.
 \label{eq:PsiBarBoundNew}
\end{equation}
The bounds \eqref{eq:PsiDerivativeNew} and $\sup_t\|\dot G(t)\|<\infty$ show that the right-hand side is uniformly bounded. This proves \eqref{eq:DeltaBbound}--\eqref{eq:Psi2bound}. The bound $\delta_B<1$ ensures that $G$ is invertible.

\section{Proof of Theorem \ref{th:demand}}
\label{app:demand}

We use the matrices in \eqref{eq:Qr}--\eqref{eq:closed2}. Since the columns of $\overline W$ form an orthonormal basis of the null space of $W$, we have $W\overline W=0$. Define $C_W=\overline W(B_0\overline W)^+$ and $F_D=W^T(WW^T)^{-1}$. By Assumption~\ref{A6}, $B_0\overline W$ has full row rank. Therefore $B_0C_W=I_m$, $WC_W=0$, and $WF_D=I_{n_r}$. Control \eqref{eq:Qr} can be written as $Q=C_W[-K_1\Sabs{\sigma}^{1/(1-l)}+\nu]+F_DD$, and direct multiplication gives $WQ=D$.

We next check that the bounds in Appendix~\ref{app:control} remain valid for the constrained control. Assumption~\ref{A1} bounds the applied $Q$ and the cumulative-flux condition $\bar u$ through \eqref{eq:diff2}. These are the input properties used in Appendices~\ref{app:control_bound_diff}--\ref{app:control_bound_pert}. The additional demand term in $\Psi_W=\Psi+BF_DD$ is treated below. Thus all pressure, SR, and regional estimates obtained in Appendix~\ref{app:control}, including \eqref{eq:PsiDerivativeNew}, apply here.

If $m_R=0$, then $G_W=(\beta_0/\beta)I_m$. Choosing $0<\beta_0<2\beta$ gives $\|G_W-I_m\|\leq\delta_{B,W}:=|\beta_0/\beta-1|<1$ directly. We therefore assume $m_R>0$ below.

Unlike \eqref{eq:Bbounds}, the demand constraint can couple the control channels, so $G_W$ need not be diagonal. Write $C_W=[C_u\ C_R]$, where $C_u$ and $C_R$ contain the columns associated with the pressure and SR outputs. Put $a=\beta_0/\beta>0$ and $\kappa_R=(\gamma_{1_0}R_0^*)^{-1}>0$. Since $B_0C_W=I_m$ and $B_u=aB_{u_0}$, the pressure part of $G_W$ is $[aI_{m_u}\ 0]$. We now examine the SR part, following Appendix~\ref{app:control_bound_coeff}.

For each well in an SR output region, define $h_j(t)=|V_j^*|^{-1}\int_{V_j^*}\gamma_1R\,dV$. As in Appendix~\ref{app:control_bound_SR}, there are constants $0<h_j^-\leq h_j(t)\leq h_j^+<\infty$. Let $\mathcal H=\prod_j[h_j^-,h_j^+]$, where the product includes the wells in SR regions. For any $h\in\mathcal H$, write $H_i(h)=N_i^{-1}\sum_{j\in\mathcal I_i}h_j$, using the well indices introduced in \eqref{eq:SRavg}. Thus $H_i(t)=H_i(h(t))$. The bounds $H_m,H_M$ of Appendix~\ref{app:control_bound_coeff} may be chosen so that $0<H_m\leq H_i(h)\leq H_M$ for every $h\in\mathcal H$.

In Appendix~\ref{app:control_bound_coeff}, the SR gain depended only on $H_i$. Under a demand constraint, different wells in the same region may have different weights in $C_W$. We therefore separate the regional average from the differences between wells. Define $E(h)\in\mathbb R^{m_R\times n}$ by
\begin{equation*}
 E_{ij}(h)=\begin{cases}
 \dfrac{h_j-H_i(h)}{\beta|V_{R_i}|},&j\in\mathcal I_i,\\
 0,&\text{otherwise}.
 \end{cases}
\end{equation*}
By \eqref{eq:B1}--\eqref{eq:B01}, the SR input matrix is $B_R(t)=\kappa_R[a\operatorname{diag}(H_i(t))B_{R_0}+E(h(t))]$. Since $B_{R_0}C_u=0$ and $B_{R_0}C_R=I_{m_R}$, we obtain
\begin{equation}
 G_W(t)=\begin{bmatrix}
 aI_{m_u}&0\\
 \kappa_R E(h(t))C_u&\kappa_R\bigl[a\operatorname{diag}(H_i(t))+E(h(t))C_R\bigr]
 \end{bmatrix}.
 \label{eq:GWblockNew}
\end{equation}
The diagonal term contains the positive regional SR sensitivities already bounded in Appendix~\ref{app:control_bound_coeff}. The terms involving $E$ describe the additional coupling caused by the demand allocation. In particular, $E(h)C_W=0$ recovers the diagonal gain of Appendix~\ref{app:control_bound_coeff}.

We first choose $\beta_0$ and $W$. Take $0<\beta_0<2\beta$ and set $\varepsilon_\beta=|a-1|<1$, as in \eqref{eq:constants}. For a proposed demand matrix $W$, check that $W$ and $B_0\overline W$ have full row rank. We also need the SR part of \eqref{eq:GWblockNew} to remain positive before multiplication by $\kappa_R$. This is precisely the condition in Assumption~\ref{A6}: there must be a constant $\mu>0$ such that
\begin{equation}
 a\operatorname{diag}(H_i(h))+
 \frac{E(h)C_R+C_R^TE(h)^T}{2}\succeq\mu I_{m_R}
 \quad\text{for all }h\in\mathcal H.
 \label{eq:WpositiveNew}
\end{equation}
This inequality can be checked directly by requiring the smallest eigenvalue of its left-hand side to be positive for every admissible $h$. Since both $H_i(h)$ and $E(h)$ are affine in $h$, it is enough to check \eqref{eq:WpositiveNew} at the vertices of $\mathcal H$. Once $W$ passes the rank and positivity tests, put
\begin{equation*}
 T(h)=[\,E(h)C_u\quad a\operatorname{diag}(H_i(h))+E(h)C_R\,],
 \qquad M=\max_{h\in\mathcal H}\|T(h)\|.
\end{equation*}
The maximum can also be evaluated at the vertices. By construction, $\|T(h)\|\leq M$ and $\|E(h)C_u\|\leq M$ for all $h\in\mathcal H$. The constants $\mu$ and $M$ are independent of $t$, with $0<\mu\leq M$. They provide a lower bound on the positive SR control action and an upper bound on its magnitude.

We can now choose $\gamma_{1_0}$ and $R_0^*$. For $x=(x_u,x_R)\in\mathbb R^m$, the block structure in \eqref{eq:GWblockNew} gives
\begin{equation*}
 \begin{split}
 \|(G_W-I_m)x\|^2={}&\varepsilon_\beta^2\|x_u\|^2+\|x_R\|^2
 +\kappa_R^2\|T(h)x\|^2\\
 &-2\kappa_R x_R^TT(h)x.
 \end{split}
\end{equation*}
Here $x_R^TT(h)x=x_R^TE(h)C_ux_u+x_R^T[a\operatorname{diag}(H_i(h))+E(h)C_R]x_R$.
By \eqref{eq:WpositiveNew}, the SR quadratic term satisfies $x_R^T[a\operatorname{diag}(H_i(h))+E(h)C_R]x_R\geq\mu\|x_R\|^2$. The remaining cross term is bounded by Young's inequality, using $\|E(h)C_u\|\leq M$, as $2|x_R^TE(h)C_ux_u|\leq\mu\|x_R\|^2
 +\frac{M^2}{\mu}\|x_u\|^2$. Consequently,
\begin{equation*}
 \begin{split}
 \|(G_W-I_m)x\|^2\leq{}&
 \left(\varepsilon_\beta^2+\frac{\kappa_RM^2}{\mu}
 +\kappa_R^2M^2\right)\|x_u\|^2\\
 &+\left(1-\kappa_R\mu+\kappa_R^2M^2\right)\|x_R\|^2.
 \end{split}
\end{equation*}
Choose the nominal SR scaling as
\begin{equation}
 (\kappa_R)^{-1}=\gamma_{1_0}R_0^*=\frac{2M^2}{(1-\varepsilon_\beta^2)\mu}.
 \label{eq:demandNominalChoiceNew}
\end{equation}
Since $\mu>0$ and $\varepsilon_\beta<1$, this choice is always positive. Substitution gives
\begin{equation}
 \|G_W(t)-I_m\|\leq
 \max\!\left\{
 \sqrt{1-\frac{1-\varepsilon_\beta^2}{4}},
 \sqrt{1-\frac{(1-\varepsilon_\beta^4)\mu^2}{4M^2}}
 \right\}=:\delta_{B,W}<1.
 \label{eq:GWboundNew}
\end{equation}
Thus the constructive order is to choose $\beta_0$, select $W$ using the rank and positivity tests, and finally select $\gamma_{1_0}R_0^*$ using \eqref{eq:demandNominalChoiceNew}. The value of $\beta_0$ controls the pressure uncertainty $\varepsilon_\beta$. Changing it rescales both terms in \eqref{eq:WpositiveNew} by the same positive factor and cannot correct an unsuitable $W$. The choice of $\gamma_{1_0}R_0^*$ then controls the magnitude of the SR gain, but not its sign. If $E(h)C_W=0$, the gain is diagonal and the less conservative midpoint choice $\gamma_{1_0}R_0^*=a(H_m+H_M)/2$ from Appendix~\ref{app:control_bound_coeff} applies. The bound \eqref{eq:GWboundNew} ensures that $G_W$ is invertible. 

Finally, we prove the bound on the normalized demand perturbation. From \eqref{eq:Psibar2}, $\Psi_W=\Psi+BF_DD$. The matrices $C_W$ and $F_D$ are constant. Since $D=WQ$, Assumption~\ref{A1} gives $\|D(t)\|\leq\|W\|L_Q$. Assumption~\ref{A6} also gives $\sup_t\|\dot D(t)\|<\infty$. The SR entries of $B$ are well averages of $\gamma_1R$, as in \eqref{eq:B1}. By the same Cauchy--Schwarz estimate used for $\dot H_i$ in A.4, $\sup_t(\|B(t)\|+\|\dot B(t)\|)<\infty$. Thus $\dot\Psi_W=\dot\Psi+\dot BF_DD+BF_D\dot D$, $\dot G_W=\dot BC_W$, and all the terms on the right are uniformly bounded. It follows from \eqref{eq:GWboundNew} (or the pressure-only bound when $m_R=0$) that $\|G_W^{-1}\|\leq(1-\delta_{B,W})^{-1}$. Differentiating the inverse as in A.4 gives
\begin{equation}
 \begin{split}
 \left\|\frac{d}{dt}(G_W^{-1}\Psi_W)\right\|
 &\leq\frac{\|\dot\Psi_W\|}{1-\delta_{B,W}}
 +\frac{\|\dot G_W\|\,\|\Psi_W\|}{(1-\delta_{B,W})^2}.
 \end{split}
 \label{eq:PsiWDotBoundNew}
\end{equation}
The right-hand side is uniformly bounded because $\Psi_W$ is bounded by \eqref{eq:PsiDerivativeNew} and $\sup_t\|B(t)\|\|F_D\|\|D(t)\|<\infty$. This establishes the bounds required in Theorem~\ref{th:demand}. The stated tracking properties follow under the gain and uncertainty conditions of \citep{b:Mathey-Moreno-2024}.

\section{Model implementation}
\label{app:sim}

For ease of implementation and visualization, the diffusion equation~\eqref{eq:diff}, with $n_d=3$ for Groningen, is approximated by a two-dimensional model obtained by averaging over the reservoir thickness (see \citep{b:Gutierrez-Stefanou-2024,doi:10.1144/SP528-2022-169} for related approaches). Let $x=(\hat{x},x_3)$, with $\hat{x}=(x_1,x_2)^\top\in\Omega\subset\mathbb{R}^2$, and let $h(\hat{x})$ denote the reservoir thickness. We define the depth-averaged pressure and horizontal flux as
\begin{equation*}
\hat{u}(\hat{x},t)=\frac{1}{h(\hat{x})}\int_0^{h(\hat{x})}u(x,t)\,dx_3,\qquad
\hat{q}(\hat{x},t)=\frac{1}{h(\hat{x})}\int_0^{h(\hat{x})}(q_1,q_2)^\top\,dx_3.
\end{equation*}
Integrating the first equation in \eqref{eq:diff} over the thickness and using the no-flux conditions on the upper and lower boundaries gives
\begin{equation*}
\hat{u}_t=-\frac{1}{\beta h}\nabla^{\mathrm{2D}}\cdot(h\hat{q})+\frac{1}{\beta h}\sum_{i=1}^n\hat{\mathcal{B}}_i(\hat{x})Q_i(t),\qquad
\hat{\mathcal{B}}_i(\hat{x})=\int_0^{h(\hat{x})}\mathcal{B}_i(x)\,dx_3,
\end{equation*}
where $\nabla^{\mathrm{2D}}$ denotes the horizontal gradient. As a reduced Darcy law, we approximate the depth-averaged horizontal flux by $\hat{q}=-\hat{K}\nabla^{\mathrm{2D}}\hat{u}$. When vertical pressure variations are neglected, the effective horizontal coefficient is given by $\hat{K}(\hat{x})=h(\hat{x})^{-1}\int_0^{h(\hat{x})}[k(x)/\eta(x)]_{\mathrm{2D}}\,dx_3$, where the subscript $\mathrm{2D}$ denotes the horizontal $2\times2$ block of the permeability-to-viscosity tensor. The reduced diffusion equation becomes
\begin{equation}
\hat{u}_t=\frac{1}{\beta h}\nabla^{\mathrm{2D}}\cdot\bigl(h\hat{K}\nabla^{\mathrm{2D}}\hat{u}\bigr)+\frac{1}{\beta h}\sum_{i=1}^n\hat{\mathcal{B}}_i(\hat{x})Q_i(t).
\label{eq:diff2D}
\end{equation}

For the two-dimensional seismicity rate, we adopt the model
\begin{equation}
\hat{R}_t(\hat{x},t)=\hat{R}(\hat{x},t)\left\{s\gamma_1(\hat{x})\hat{u}_t(\hat{x},t)-\gamma_2(\hat{x},t)\left[\hat{R}(\hat{x},t)-\hat{R}^*(\hat{x})\right]\right\}.
\label{eq:SR2D}
\end{equation}

We briefly check the properties needed to apply the analysis of Appendix~\ref{app:control}. Let $\Omega$ be bounded, connected and Lipschitz, and let $h\in W^{1,\infty}(\Omega)$ satisfy $0<h_m\le h(\hat{x})\le h_M<\infty$. The effective coefficient $\hat{K}$ is symmetric and uniformly positive definite under the bounds in Assumption~\ref{A3}. We consider the corresponding local regularity and well-region conditions of Assumptions~\ref{A3}--\ref{A5}. Define $A_h=-\beta^{-1}h^{-1}\nabla^{\mathrm{2D}}\cdot(h\hat{K}\nabla^{\mathrm{2D}}\cdot)$ in $H=L^2(\Omega,h\,d\hat{x})$, with homogeneous Neumann conditions. Its weighted inner product satisfies
\begin{equation*}
\langle v,A_hv\rangle_H=\frac{1}{\beta}\int_\Omega h(\nabla^{\mathrm{2D}}v)^\top\hat{K}\nabla^{\mathrm{2D}}v\,d\hat{x}\ge0.
\end{equation*}
Hence $A_h$ is nonnegative and self-adjoint, with constants as its kernel. The weighted Poincar\'e--Wirtinger inequality gives a positive spectral gap on the weighted mean-zero subspace. Its resolvent is compact and its analytic heat semigroup satisfies the same type of off-diagonal estimates used in Appendix~\ref{app:control}. In two dimensions, the bound $\|e^{-tA_h}\|_{H\to L^\infty}\le C(1+t^{-1/2})$ gives $D(A_h)\hookrightarrow L^\infty(\Omega)$. The local regularity of $\hat{K}$ also gives the required interior $H^2$ estimates \citep{b:EvansPDE,b:Pazy,b:Ouhabaz2005}.

The estimates in Appendix~\ref{app:control} therefore follow in the weighted space, using the mean $\int_\Omega h\hat{u}\,d\hat{x}/\int_\Omega h\,d\hat{x}$. The local regularity and off-diagonal bounds allow the regional perturbation derivatives to be bounded as in Appendix~\ref{app:control_bound_pert}. For a two-dimensional output region $\Omega_i$, we use the volume average $\int_{\Omega_i}h\hat{u}\,d\hat{x}/\int_{\Omega_i}h\,d\hat{x}$, and the same definition for $\hat{R}$. Moreover, the reduced well coefficients $\hat{\mathcal{B}}_i/h$ satisfy $\int_\Omega h(\hat{\mathcal{B}}_i/h)\,d\hat{x}=1$, preserving the well normalization used in the control design. With the corresponding assumptions on the finite-support wells and the input matrices, the bounds of Appendix~\ref{app:control_bound_coeff} and Appendix~\ref{app:demand} follow in the same way. Thus, the qualitative tracking properties of Theorems~\ref{th:1} and~\ref{th:demand} apply to the reduced model under the same control conditions.

Equation~\eqref{eq:SR2D} is not, in general, the direct depth average of \eqref{eq:SR}, since averaging its nonlinear terms introduces additional terms involving the vertical variations of the SR, pressure rate and model coefficients. Here, \eqref{eq:SR2D} is used as a simpler two-dimensional model to implement and visualize the control. It retains the pressure-rate dependence and the nonlinear evolution of the SR considered in \citep{Dieterich1994,Segall2015}. Reduced models relating reservoir operations and stress changes to induced seismicity have also been used for Groningen \citep{b:https://doi.org/10.1029/2020JB020013,doi:10.1144/SP528-2022-169,b:doi.org/10.1029/2023GL105455,BourneOates2017}. In Section~\ref{sec:simulations}, its parameters are fitted using the Groningen seismicity data. The resulting model reproduces the spatial distribution of the seismic events and the evolution of the average SR and cumulative events, as shown in Figs.~\ref{fig:density} and~\ref{fig:validation}. These results support its use in the present two-dimensional simulations without implying that it is equivalent to the full three-dimensional SR equation.

\begin{figure}[ht!]
  \centering
  \includegraphics[width=5.0cm,keepaspectratio]{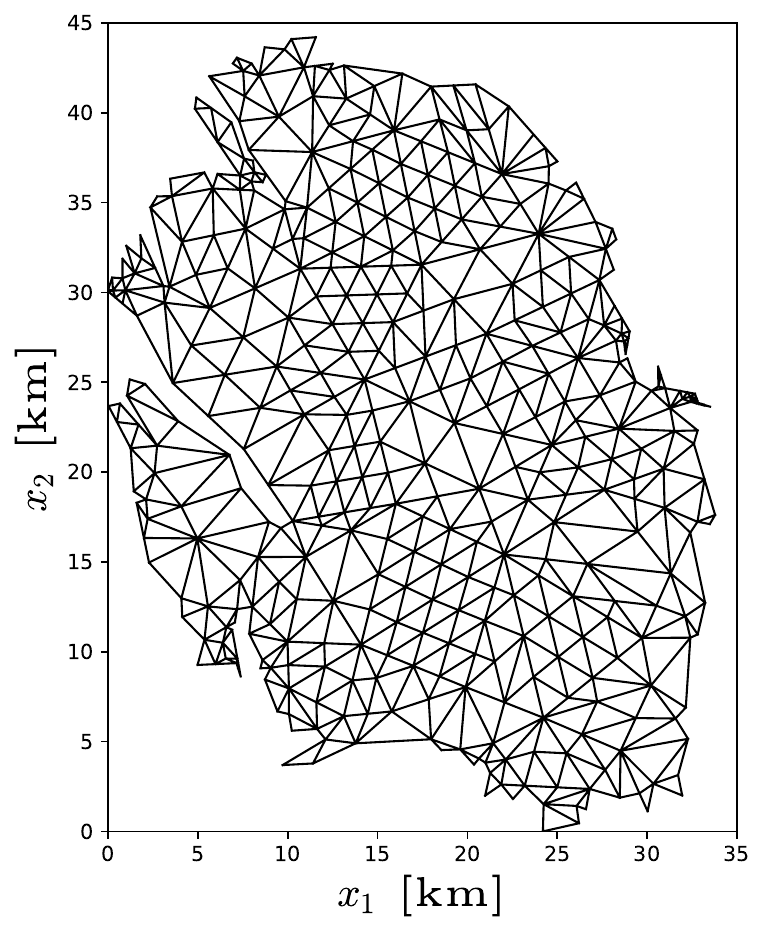}
  \caption{Discretization of the reservoir in 446 elements using triangular mesh.}
  \label{fig:Groningen_dis}
\end{figure}

The two-dimensional diffusion equation \eqref{eq:diff2D} is numerically solved in Section~\ref{sec:simulations} using first-order finite elements~\citep{b:skfem2020}. A triangular mesh with 446 elements was employed (see Fig.~\ref{fig:Groningen_dis}). To assess the accuracy of the spatial discretization, we conducted a mesh convergence study of the solution, $u(x,t)$, and the error vector, $\sigma(t)$, with respect to the average normalized mesh size, denoted by $\overline{h}$. 

\begin{figure}[ht!]
  \centering
  \includegraphics[width=6.0cm,keepaspectratio]{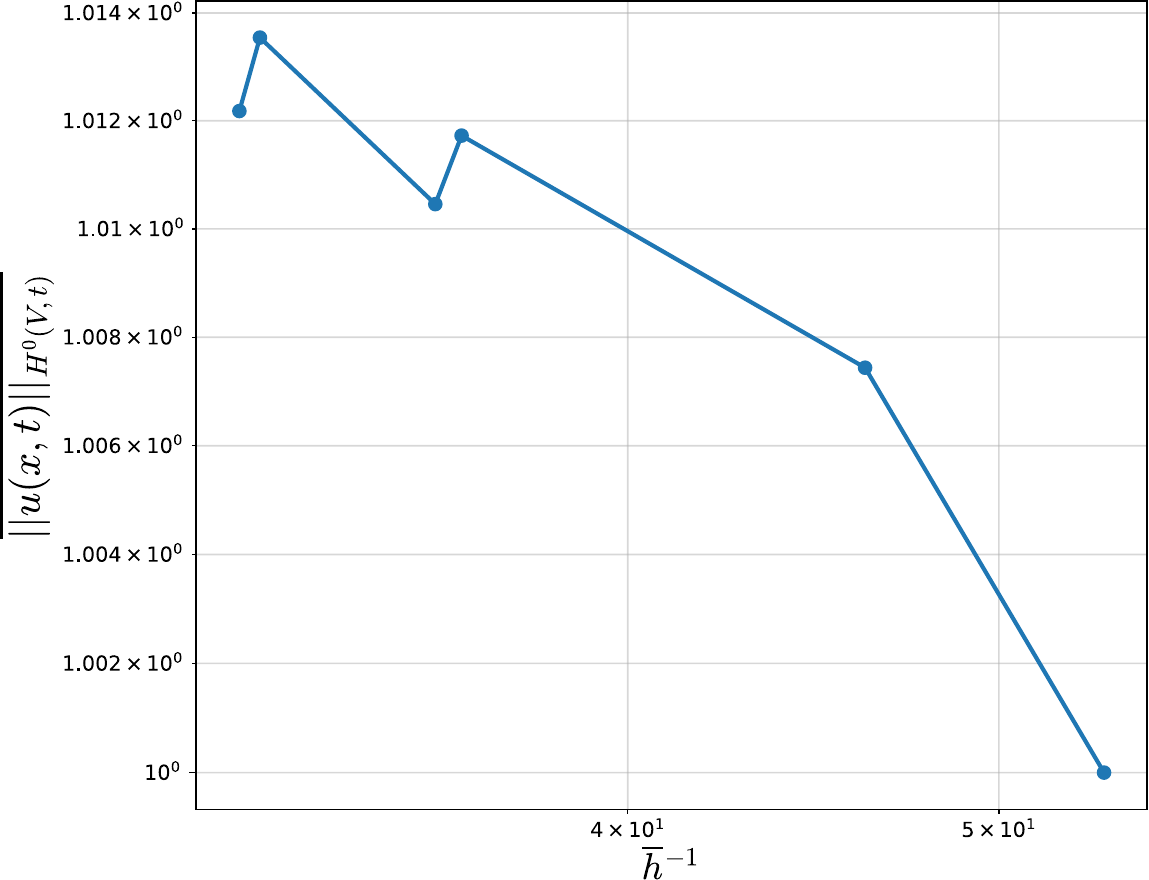}
  \includegraphics[width=5.6cm,keepaspectratio]{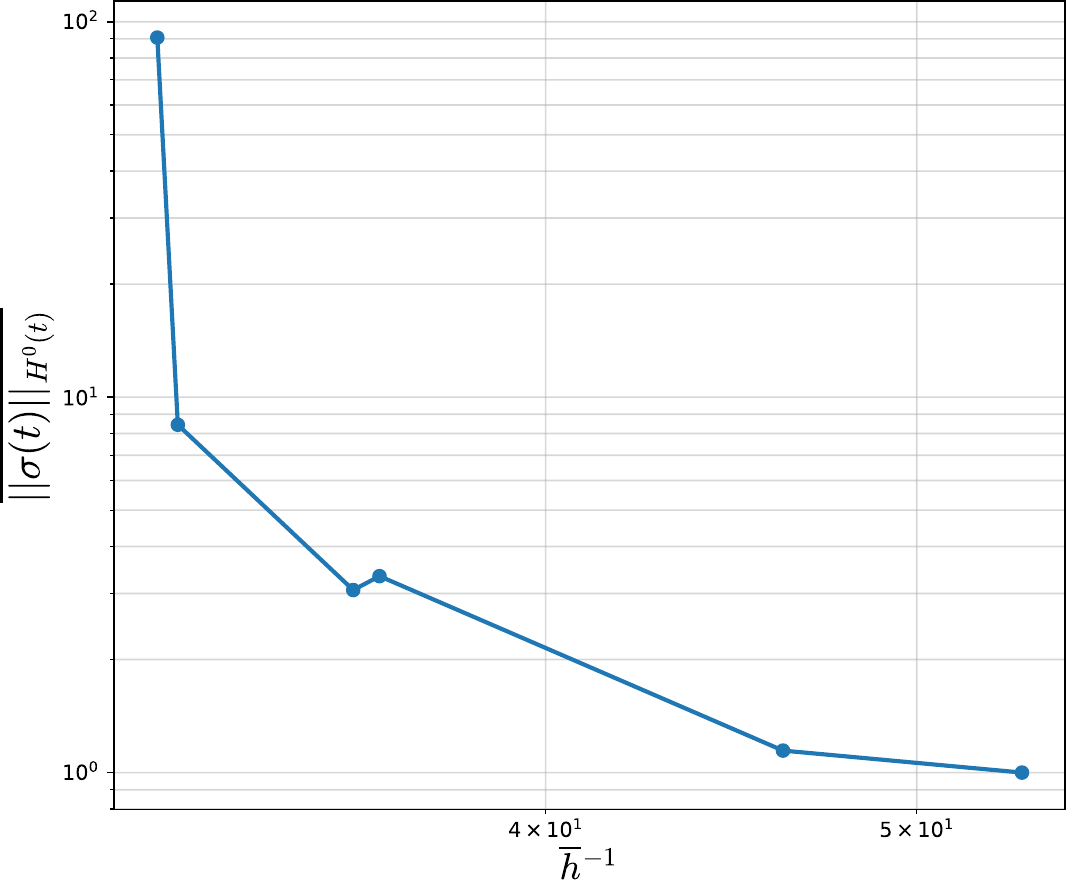}
  \caption{Mesh convergence analysis over $u(x,t)$ and $\sigma(t)$.}
  \label{fig:convergence}
\end{figure}

Figure~\ref{fig:convergence} shows the convergence behaviour of the normalized $L^2$-norms of the solution and error, \textit{i.e.}, $\overline{\|u(x,t)\|_{L^2(V,t)}} := \frac{\sqrt{\int \|u(x,t)\|^2_{L^2(V)} \, dt}}{u_{\mathrm{norm}}}$, $\overline{\|\sigma(t)\|_{L^2(t)}} := \frac{\sqrt{\int \|\sigma(t)\|^2 \, dt}}{\sigma_{\mathrm{norm}}}$, where $u_{\mathrm{norm}}$ and $\sigma_{\mathrm{norm}}$ denote the corresponding norms computed using the finest mesh. These results confirm that the error converges as the mesh is refined, indicating spatial convergence.

Finally, both systems \eqref{eq:diff2D}--\eqref{eq:SR2D} were discretized in time using the implicit backward differentiation formula (BDF) method implemented in Python~\citep{b:doi:10.1137/S1064827594276424}. The simulations start in December 1991, when seismic events began to be recorded in Groningen, using the pressure and SR states obtained from the preceding historical simulation, initialized with a zero pressure difference field and the positive background SR. The controller is initialized at the same historical operating point, with $Q(0)=Q_b$, where $Q_b$ denotes the historical well fluxes at that date. The historical well fluxes are continuous and piecewise differentiable, with bounded derivatives. Together with the semigroup estimates of Appendix \ref{app:control_bound_diff}, this initialization was verified to satisfy $p(0)\in D(A)$ and $p_t(0)\in D(A)$ in the shifted system. Furthermore, the positivity and local regularity results of Appendix \ref{app:control_bound_SR} ensure that the initial SR field is bounded, strictly positive near the wells, and belongs to $H^2(U_j)$ under the spatial regularity conditions of Assumption \ref{A3}. The neighborhoods $U_j$ are those specified in Assumption \ref{A3}, consistently with the separation condition in Assumption \ref{A4}. Hence, all the initial conditions in Assumption \ref{A5} are satisfied.

\end{appendices}

\clearpage
\section*{Graphical Abstract}
\begin{figure}[h!]
    \centering
    \includegraphics[width=\textwidth]{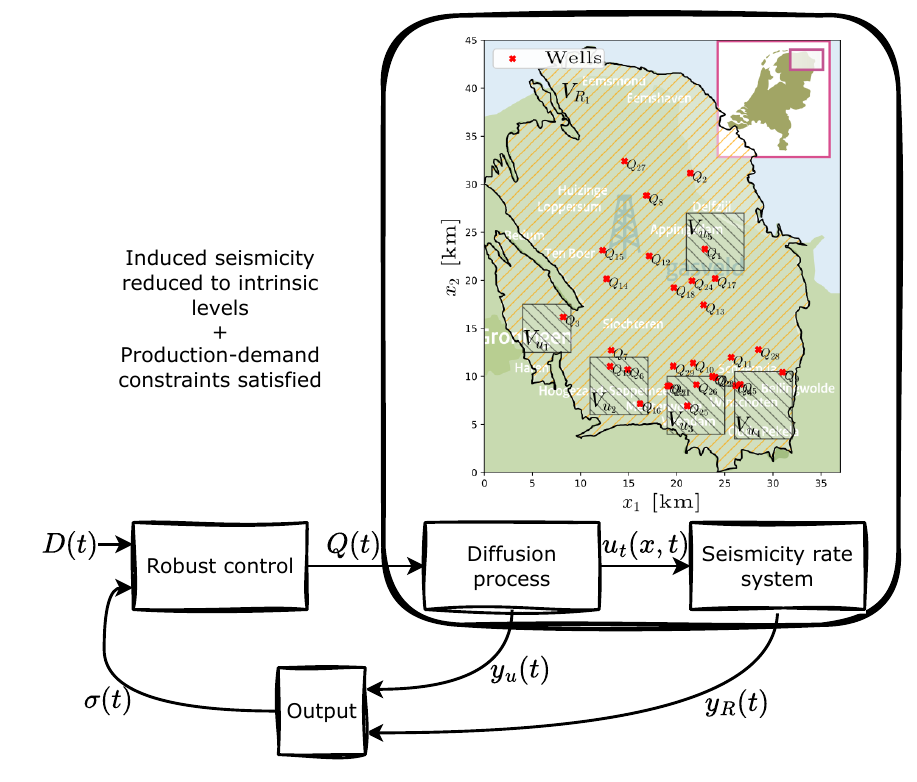}
\end{figure}


\bibliography{Bibliografias}

@article{BourneOates2017,
  author  = {Bourne, S. J. and Oates, S. J.},
  title   = {Extreme Threshold Failures Within a Heterogeneous Elastic Thin Sheet and the Spatial-Temporal Development of Induced Seismicity Within the Groningen Gas Field},
  journal = {Journal of Geophysical Research: Solid Earth},
  year    = {2017},
  volume  = {122},
  number  = {12},
  pages   = {10299--10320},
  doi     = {10.1002/2017JB014356}
}

@article{Segall1989,
  author  = {Segall, Paul},
  title   = {Earthquakes Triggered by Fluid Extraction},
  journal = {Geology},
  year    = {1989},
  volume  = {17},
  number  = {10},
  pages   = {942--946},
  doi     = {10.1130/0091-7613(1989)017<0942:ETBFE>2.3.CO;2}
}

@article{Smith2020,
  author  = {Smith, Jonathan D. and White, Robert S. and Avouac, Jean-Philippe and Bourne, Stephen},
  title   = {Probabilistic Earthquake Locations of Induced Seismicity in the Groningen Region, the Netherlands},
  journal = {Geophysical Journal International},
  year    = {2020},
  volume  = {222},
  number  = {1},
  pages   = {507--516},
  doi     = {10.1093/gji/ggaa179}
}

@article{DempseySuckale2017,
  author  = {Dempsey, David and Suckale, Jenny},
  title   = {Physics-Based Forecasting of Induced Seismicity at Groningen Gas Field, the Netherlands},
  journal = {Geophysical Research Letters},
  year    = {2017},
  volume  = {44},
  number  = {15},
  pages   = {7773--7782},
  doi     = {10.1002/2017GL073878}
}

@article{Candela2018,
  author  = {Candela, Thibault and Wassing, Brecht and {ter Heege}, Jan and Buijze, Loes},
  title   = {How Earthquakes Are Induced},
  journal = {Science},
  year    = {2018},
  volume  = {360},
  number  = {6389},
  pages   = {598--600},
  doi     = {10.1126/science.aat2776}
}

@unpublished{b:Gutierrez-Stefanou-2026_c,
  author = {Diego Gutiérrez-Oribio and Ioannis Stefanou},
  title = {{Modular Sign Compensation for MIMO Systems with Unknown Control Direction: An Exact Nominal Recovery Approach}},
  year = {2026},
  note = {\href{https://doi.org/10.48550/arXiv.2607.14839
}{DOI:10.48550/arXiv.2607.14839}}
}

@article{Candela2019Groningen,
  author  = {Candela, Thibault and Osinga, Sander and Ampuero, Jean-Paul and Wassing, Brecht and Pluymaekers, Maarten and Fokker, Peter A. and van Wees, Jan-Diederik and de Waal, Hans A. and Muntendam-Bos, Annemarie G.},
  title   = {Depletion-Induced Seismicity at the Groningen Gas Field: Coulomb Rate-and-State Models Including Differential Compaction Effect},
  journal = {Journal of Geophysical Research: Solid Earth},
  volume  = {124},
  number  = {7},
  pages   = {7081--7104},
  year    = {2019},
  doi     = {10.1029/2018JB016670}
}

@article{b:Ji2023,
  author  = {Chunting Ji and Zhengqiang Zhang},
  title   = {Adaptive observer for {ODE}-{PDE} cascade systems subject to nonlinear dynamics and uncertain parameters},
  journal = {Nonlinear Dynamics},
  year    = {2023},
  volume  = {111},
  number  = {18},
  pages   = {17317--17336},
  doi     = {10.1007/s11071-023-08780-5},
}

@article{b:deBuhanFrey2011,
  author  = {Maya de Buhan and Pascal Frey},
  title   = {A generalized model of nonlinear viscoelasticity: numerical issues and applications},
  journal = {International Journal for Numerical Methods in Engineering},
  year    = {2011},
  volume  = {86},
  number  = {12},
  pages   = {1544--1557}
}

@article{b:Rizzi2021,
  author  = {Gianluca Rizzi and Mohamed Collet and Francesco Ghiba and Angela Madeo and Patrizio Neff},
  title   = {Exploring Metamaterials' Structures Through the Relaxed Micromorphic Model: Switching an Acoustic Screen Into an Acoustic Absorber},
  journal = {Frontiers in Materials},
  year    = {2021},
  volume  = {7},
  pages   = {589701},
  doi     = {10.3389/fmats.2020.589701}
}

@techreport{b:NAM2016GPM,
  author      = {{NAM}},
  title       = {Groningen Pressure Maintenance (GPM) Study, Progress Report July 2016},
  institution = {Nederlandse Aardolie Maatschappij B.V.},
  year        = {2016},
  address     = {Assen, The Netherlands},
  note        = {Progress report},
}

@article {b:10.7554/eLife.37836,
title = {Competition for synaptic building blocks shapes synaptic plasticity},
author = {Triesch, Jochen and Vo, Anh Duong and Hafner, Anne-Sophie},
volume = 7,
year = 2018,
month = {sep},
pub_date = {2018-09-17},
pages = {e37836},
doi = {10.7554/eLife.37836},
journal = {eLife},
publisher = {eLife Sciences Publications, Ltd},
}

@article{b:GALLAY2022103387,
title = {Propagation fronts in a simplified model of tumor growth with degenerate cross-dependent self-diffusivity},
journal = {Nonlinear Analysis: Real World Applications},
volume = {63},
pages = {103387},
year = {2022},
doi = {https://doi.org/10.1016/j.nonrwa.2021.103387},
author = {Thierry Gallay and Corrado Mascia}
}

@book{b:helander2005collisional,
  title={Collisional Transport in Magnetized Plasmas},
  author={Helander, P. and Sigmar, D.J.},
  series={Cambridge Monographs on Plasma Physics},
  address={Cambridge, U.K.},
  year={2005},
  publisher={Cambridge University Press}
}

@article{b:KURDYUMOV2015981,
title = {Structure and stability of premixed flames stabilized behind the trailing edge of a cylindrical rod at low Lewis numbers},
journal = {Proceedings of the Combustion Institute},
volume = {35},
number = {1},
pages = {981-988},
year = {2015},
doi = {https://doi.org/10.1016/j.proci.2014.05.056},
author = {V.N. Kurdyumov and Y.L. Shoshin and L.P.H. {de Goey}}
}

@ARTICLE{b:7307993,
  author={Deutscher, Joachim},
  journal={IEEE Transactions on Automatic Control}, 
  title={{Backstepping Design of Robust Output Feedback Regulators for Boundary Controlled Parabolic PDEs}}, 
  year={2016},
  volume={61},
  number={8},
  pages={2288-2294},
  doi={10.1109/TAC.2015.2491718}
}

@book{b:martinez2018optimal,
  title={{Optimal control of PDEs under uncertainty: an introduction with application to optimal shape design of structures}},
  author={Mart{\'\i}nez-Frutos, Jes{\'u}s and Esparza, Francisco Periago},
  year={2018},
  publisher={Springer},
  address={Switzerland}
}

@book{b:fursikov1999optimal,
  title={{Optimal control of distributed systems. Theory and applications}},
  author={Fursikov, A. V.},
  year={1999},
  publisher={American Mathematical Society},
  address={Rhode Island, USA}
}

@article{b:https://doi.org/10.1002/nag.3923,
author = {Gutiérrez-Oribio, Diego and Stathas, Alexandros and Stefanou, Ioannis},
title = {{AI-Driven Approach for Sustainable Extraction of Earth's Subsurface Renewable Energy While Minimizing Seismic Activity}},
journal = {International Journal for Numerical and Analytical Methods in Geomechanics},
volume = {49},
number = {4},
pages = {1126-1138},
year = {2025},
doi = {https://doi.org/10.1002/nag.3923}
}

@article{b:kim2023,
  author = {Kim, T. and Avouac, J.-P.},
  title = {{Stress-Based and Convolutional Forecasting of Injection-Induced Seismicity: Application to the Otaniemi Geothermal Reservoir Stimulation}},
  journal = {Journal of Geophysical Research: Solid Earth},
  year = {2023},
  doi = {10.1029/2023JB024271}
}

@article{https://doi.org/10.1029/2018WR023587,
author = {Dempsey, David and Riffault, Jeremy},
title = {{Response of Induced Seismicity to Injection Rate Reduction: Models of Delay, Decay, Quiescence, Recovery, and Oklahoma}},
journal = {Water Resources Research},
volume = {55},
number = {1},
pages = {656-681},
doi = {https://doi.org/10.1029/2018WR023587},
year = {2019}
}

@article{b:doi:10.1137/S1064827594276424,
author = {Shampine, Lawrence F. and Reichelt, Mark W.},
title = {{The MATLAB ODE Suite}},
journal = {SIAM Journal on Scientific Computing},
volume = {18},
number = {1},
pages = {1-22},
year = {1997},
doi = {10.1137/S1064827594276424}
}

@article{b:skfem2020,
  doi = {10.21105/joss.02369},
  year = {2020},
  volume = {5},
  number = {52},
  pages = {2369},
  author = {Tom Gustafsson and G. D. McBain},
  title = {scikit-fem: A {P}ython package for finite element assembly},
  journal = {Journal of Open Source Software}
}

@article{b:https://doi.org/10.1029/2024GL110139,
author = {Tamama, Y. and Acosta, M. and Bourne, S. J. and Avouac, J. P.},
title = {{Earthquake Growth Inhibited at Higher Coulomb Stress Change Rate at Groningen}},
journal = {Geophysical Research Letters},
volume = {51},
number = {20},
pages = {e2024GL110139},
doi = {https://doi.org/10.1029/2024GL110139},
year = {2024}
}

@article{b:https://doi.org/10.1029/2020JB020013,
author = {Bourne, S. J. and Oates, S. J.},
title = {{Stress-Dependent Magnitudes of Induced Earthquakes in the Groningen Gas Field}},
journal = {Journal of Geophysical Research: Solid Earth},
volume = {125},
number = {11},
pages = {e2020JB020013},
doi = {https://doi.org/10.1029/2020JB020013},
year = {2020}
}

@article{b:10.1785/0220230179,
    author = {Kaveh, Hojjat and Batlle, Pau and Acosta, Mateo and Kulkarni, Pranav and Bourne, Stephen J. and Avouac, Jean Philippe},
    title = {{Induced Seismicity Forecasting with Uncertainty Quantification: Application to the Groningen Gas Field}},
    journal = {Seismological Research Letters},
    volume = {95},
    number = {2A},
    pages = {773-790},
    year = {2023},
    month = {12},
    doi = {10.1785/0220230179},
}

@article{b:zhao2021backstepping,
  title={{Backstepping-based decentralized fault-tolerant control of hypersonic vehicles in PDE-ODE form}},
  author={Zhao, Dong and Jiang, Bin and Yang, Hao},
  journal={IEEE Transactions on Automatic Control},
  volume={67},
  number={3},
  pages={1210--1225},
  year={2021},
  publisher={IEEE}
}

@article{b:wang2015sliding,
  title={{Sliding mode control to stabilization of cascaded heat PDE--ODE systems subject to boundary control matched disturbance}},
  author={Wang, Jun-Min and Liu, Jun-Jun and Ren, Beibei and Chen, Jinhao},
  journal={Automatica},
  volume={52},
  pages={23--34},
  year={2015},
  publisher={Elsevier}
}

@article{b:tang2011state,
  title={{State and output feedback boundary control for a coupled PDE--ODE system}},
  author={Tang, Shuxia and Xie, Chengkang},
  journal={Systems \& Control Letters},
  volume={60},
  number={8},
  pages={540--545},
  year={2011},
  publisher={Elsevier}
}

@article{b:deutscher2020output,
  title={{Output Feedback Control of Coupled Linear Parabolic ODE--PDE--ODE Systems}},
  author={Deutscher, Joachim and Gehring, Nicole},
  journal={IEEE Transactions on Automatic Control},
  volume={66},
  number={10},
  pages={4668--4683},
  year={2020},
  publisher={IEEE}
}

@article{b:susto2010control,
  title={{Control of PDE--ODE cascades with Neumann interconnections}},
  author={Susto, Gian Antonio and Krstic, Miroslav},
  journal={Journal of the Franklin Institute},
  volume={347},
  number={1},
  pages={284--314},
  year={2010},
  publisher={Elsevier}
}

@article{b:quarteroni2003analysis,
  title={{Analysis of a geometrical multiscale model based on the coupling of ODE and PDE for blood flow simulations}},
  author={Quarteroni, Alfio and Veneziani, Alessandro},
  journal={Multiscale Modeling \& Simulation},
  volume={1},
  number={2},
  pages={173--195},
  year={2003},
  publisher={SIAM}
}

@article{b:koch2022sliding,
  title={{Sliding-Mode Boundary Control of an In-Line Heating System Governed by Coupled PDE/ODE Dynamics}},
  author={Koch, Stefan and Pilloni, Alessandro and Pisano, Alessandro and Usai, Elio},
  journal={IEEE Transactions on Control Systems Technology},
  volume={30},
  number={6},
  pages={2689-2697},
  year={2022},
  publisher={IEEE}
}

@article{b:zhong2016coupled,
  title={{Coupled electromagnetic and heat transfer ODE model for microwave heating with temperature-dependent permittivity}},
  author={Zhong, Jiaqi and Liang, Shan and Yuan, Yupeng and Xiong, Qingyu},
  journal={IEEE Transactions on Microwave Theory and Techniques},
  volume={64},
  number={8},
  pages={2467-2477},
  year={2016},
  publisher={IEEE}
}

@BOOK{b:EvansPDE,
    Address        = {Rhode Island, USA},
    Author         = {L. C. Evans},
    doi            = {10.1090/gsm/019},
    Publisher      = {American Mathematical Society },
    Title          = {Partial differential equations },
    Year           = {2010}
}

@article{https://doi.org/10.1002/jgrb.50264,
author = {Shapiro, Serge A. and Krüger, Oliver S. and Dinske, Carsten},
title = {Probability of inducing given-magnitude earthquakes by perturbing finite volumes of rocks},
journal = {Journal of Geophysical Research: Solid Earth},
volume = {118},
number = {7},
pages = {3557-3575},
doi = {https://doi.org/10.1002/jgrb.50264},
year = {2013}
}

@article{doi:10.1144/SP528-2022-169,
author = {Hadrien Meyer and Jonathan D. Smith and Stephen Bourne and Jean-Philippe Avouac},
title = {An integrated framework for surface deformation modelling and induced seismicity forecasting due to reservoir operations},
journal = {Geological Society, London, Special Publications},
volume = {528},
number = {1},
pages = {299-318},
year = {2023},
doi = {10.1144/SP528-2022-169},
}

@misc{b:acosta_2023_8329298,
  author       = {Acosta, Mateo},
  title        = {{Data and software for: Acosta et al., 2023}},
  year         = 2023,
  publisher    = {Zenodo},
  doi          = {10.5281/zenodo.8329298},
  note          = {\url{https://doi.org/10.5281/zenodo.8329298}}
}

@misc{b:Groningen1,
  title        = {{Groningen gas field}},
  author       = {The NLOG Dutch Oil and Gas Portal},
  year         = 2023,
  note          = {\url{https://www.nlog.nl/groningen-gasveld}}
}

@misc{b:Groningen2,
  title        = {{Dashboard Groningen}},
  author       = {Rijksoverheid},
  year         = 2023,
  note          = {\url{https://dashboardgroningen.nl}}
}

@article{b:SMITH2022117697,
title = {{Stress-based forecasting of induced seismicity with instantaneous earthquake failure functions: Applications to the Groningen gas reservoir}},
journal = {Earth and Planetary Science Letters},
volume = {594},
pages = {117697},
year = {2022},
doi = {https://doi.org/10.1016/j.epsl.2022.117697},
author = {Jonathan D. Smith and Elías R. Heimisson and Stephen J. Bourne and Jean-Philippe Avouac},
}

@article{b:doi.org/10.1029/2023GL105455,
author = {Acosta, Mateo and Avouac, Jean-Philippe and Smith, Jonathan D. and Sirorattanakul, Krittanon and Kaveh, Hojjat and Bourne, Stephen J.},
title = {{Earthquake Nucleation Characteristics Revealed by Seismicity Response to Seasonal Stress Variations Induced by Gas Production at Groningen}},
journal = {Geophysical Research Letters},
volume = {50},
number = {19},
pages = {e2023GL105455},
doi = {https://doi.org/10.1029/2023GL105455},
year = {2023}
}

@article{b:Gutierrez-Stefanou-2024,
title = {Insights of using control theory for minimizing induced seismicity in underground reservoirs},
journal = {Geomechanics for Energy and the Environment},
pages = {100570},
year = {2024},
issn = {2352-3808},
doi = {10.1016/j.gete.2024.100570},
author = {Diego Gutiérrez-Oribio and Ioannis Stefanou},
}

@article{b:Verdon-Bommer-2021,
    author = {James P. Verdon and Julian J. Bommer },
    title = {{Green, yellow, red, or out of the blue? An assessment of Traffic Light Schemes to mitigate the impact of hydraulic fracturing-induced seismicity}},
    journal = {Journal of Seismology },
    volume = {25},
    number = {},
    pages = {301-326},
    year = {2021},
    doi = {10.1007/s10950-020-09966-9}
}

@article{b:10.1785/0220180337,
    author = {Baisch, Stefan and Koch, Christopher and Muntendam‐Bos, Annemarie},
    title = {{Traffic Light Systems: To What Extent Can Induced Seismicity Be Controlled?}},
    journal = {Seismological Research Letters},
    volume = {90},
    number = {3},
    pages = {1145-1154},
    year = {2019},
    doi = {10.1785/0220180337}
}

@article{b:10.1093/gji/ggac416,
    author = {Ji, Yinlin and Zhang, Wei and Hofmann, Hannes and Chen, Yuedu and Kluge, Christian and Zang, Arno and Zimmermann, Günter},
    title = {{Modelling of fluid pressure migration in a pressure sensitive fault zone subject to cyclic injection and implications for injection-induced seismicity}},
    journal = {Geophysical Journal International},
    volume = {232},
    number = {3},
    pages = {1655-1667},
    year = {2022},
    doi = {10.1093/gji/ggac416}
}

@article{https://doi.org/10.1007/s00603-018-1467-4,
author = {Arno Zang and Gunter Zimmermann and Hannes Hofmann and Ove Stephansson and Ki-Bok Min and Kwang Yeom Kim},
title = {{How to Reduce Fluid-Injection-Induced Seismicity}},
journal = {Rock Mechanics and Rock Engineering},
volume = {52},
number = {},
pages = {475-493},
doi = {10.1007/s00603-018-1467-4},
year = {2019}
}

@article{Dieterich1994,
	title = {A constitutive law for rate of earthquake production and its application to earthquake clustering},
	volume = {99},
	issn = {01480227},
	doi = {10.1029/93JB02581},
	number = {B2},
	journal = {Journal of Geophysical Research},
	author = {Dieterich, James H.},
	year = {1994},
	pages = {2601--2618}
}

@article{Segall2015,
	title = {Injection-induced seismicity: {Poroelastic} and earthquake nucleation effects},
	volume = {120},
	doi = {10.1002/2015JB012060},
	number = {7},
	journal = {Journal of Geophysical Research: Solid Earth},
	author = {Segall, Paul and Lu, S},
	month = jul,
	year = {2015},
	pages = {5082--5103}
}

@article{Zienkiewicz1980,
	title = {Drained, undrained, consolidating and dynamic behaviour assumptions in soils},
	volume = {30},
	issn = {17517656},
	doi = {10.1680/geot.1980.30.4.385},
	number = {4},
	journal = {Geotechnique},
	author = {Zienkiewicz, O. C. and Chang, C. T. and Bettess, P.},
	year = {1980},
	pages = {385--395}
}

@article{b:10.1785/gssrl.80.5.784,
    author = {Deichmann, Nicholas and Giardini, Domenico},
    title = {{Earthquakes Induced by the Stimulation of an Enhanced Geothermal System below Basel (Switzerland)}},
    journal = {Seismological Research Letters},
    volume = {80},
    number = {5},
    pages = {784-798},
    year = {2009},
    month = {09},
    doi = {10.1785/gssrl.80.5.784},
}

@article{b:Stey_LeMonde,
author = {Nathalie Stey},
title = {{En Alsace, les projets de géothermie profonde à l’arrêt}},
journal = {Le Monde},
year = {11/12/2020}
}

@article{b:10.1002/2016RG000542,
author = {Grigoli, Francesco and Cesca, Simone and Priolo, Enrico and Rinaldi, Antonio Pio and Clinton, John F. and Stabile, Tony A. and Dost, Bernard and Fernandez, Mariano Garcia and Wiemer, Stefan and Dahm, Torsten},
title = {Current challenges in monitoring, discrimination, and management of induced seismicity related to underground industrial activities: A European perspective},
journal = {Reviews of Geophysics},
volume = {55},
number = {2},
pages = {310-340},
doi = {10.1002/2016RG000542},
year = {2017}
}

@article{b:10.1785/0220170112,
    author = {Wilson, M. P. and Foulger, G. R. and Gluyas, J. G. and Davies, R. J. and Julian, B. R.},
    title = "{HiQuake: The Human‐Induced Earthquake Database}",
    journal = {Seismological Research Letters},
    volume = {88},
    number = {6},
    pages = {1560-1565},
    year = {2017},
    month = {10},
    doi = {10.1785/0220170112}
}

@BOOK{b:Pazy,
    Address        = {New York, USA},
    Author         = {A. Pazy},
    doi            = {10.1007/978-1-4612-5561-1},
    Publisher      = {Springer New York},
    Title          = {Semigroups of Linear Operators and Applications to Partial Differential Equations},
    Year           = {1992}
}

@article{b:Mathey-Moreno-2024,
title = {MIMO Super-Twisting Controller using a passivity-based design},
journal = {Journal of the Franklin Institute},
volume = {361},
number = {17},
pages = {107296},
year = {2024},
doi = {https://doi.org/10.1016/j.jfranklin.2024.107296},
author = {Jaime A. Moreno and Juan F. García-Mathey}
}

@article{b:doi.org/10.1029/2020GL090648,
author = {Frash, L. P. and Fu, P. and Morris, J. and Gutierrez, M. and Neupane, G. and Hampton, J. and Welch, N. J. and Carey, J. W. and Kneafsey, T.},
title = {Fracture Caging to Limit Induced Seismicity},
journal = {Geophysical Research Letters},
volume = {48},
number = {1},
pages = {e2020GL090648},
doi = {https://doi.org/10.1029/2020GL090648},
year = {2021}
}

@article{b:10.1093/gji/ggz058,
    author = {Hofmann, Hannes and Zimmermann, Günter and Farkas, Marton and Huenges, Ernst and Zang, Arno and Leonhardt, Maria and Kwiatek, Grzegorz and Martinez-Garzon, Patricia and Bohnhoff, Marco and Min, Ki-Bok and Fokker, Peter and Westaway, Rob and Bethmann, Falko and Meier, Peter and Yoon, Kern Shin and Choi, Jai Won and Lee, Tae Jong and Kim, Kwang Yeom},
    title = "{First field application of cyclic soft stimulation at the Pohang Enhanced Geothermal System site in Korea}",
    journal = {Geophysical Journal International},
    volume = {217},
    number = {2},
    pages = {926-949},
    year = {2019},
    doi = {10.1093/gji/ggz058}
}

@ARTICLE{b:Gutierrez-Orlov-Stefanou-Plestan-2023,
    Author         = {D. Guti\'errez-Oribio and Yury Orlov and I. Stefanou and F. Plestan},
    Journal        = {Systems \& Control Letters},
    Volume         = {178},
    Number         = {},
    Pages          = {105571},
    Title          = {{Robust Boundary Tracking Control of Wave PDE: Insight on Forcing Slow-Aseismic Response}},
    Year           = {2023} ,
    doi            = {10.1016/j.sysconle.2023.105571}
}

@article{b:Gutierrez-Stefanou-Plestan-2024,
    Author         = {D. Guti\'errez-Oribio and I. Stefanou and F. Plestan},
    journal={Automatica},
    Title          = {Passivity-based control of underactuated mechanical systems with Coulomb friction: Application to earthquake prevention},
    Year           = {2024},
    volume={165},
    number={111661},
    doi = {10.1016/j.automatica.2024.111661}
}

@article{b:Gutierrez-Tzortzopoulos-Stefanou-Plestan-2022,
  author={Gutiérrez-Oribio, Diego and Tzortzopoulos, Georgios and Stefanou, Ioannis and Plestan, Franck},
  journal={IEEE Transactions on Control Systems Technology}, 
  title={{Earthquake Control: An Emerging Application for Robust Control. Theory and Experimental Tests}}, 
  year={2023},
  volume={31},
  number={4},
  pages={1747-1761},
  doi={10.1109/TCST.2023.3242431}
}

@INPROCEEDINGS{b:Gutierrez-Orlov-Plestan-Stefanou-VSS2022,
    Address        = {Rio de Janeiro, Brasil},
    Author         = {D. Guti\'errez-Oribio and Y. Orlov and I. Stefanou and F. Plestan},
    Booktitle      = {{16th International Workshop on Variable Structure Systems and Sliding Mode Control}},
    Title          = {{Advances in Sliding Mode Control of Earthquakes via Boundary Tracking of Wave and Heat PDEs}},
    Year           = {2022},
    doi            = {10.1109/VSS57184.2022.9902111}
}

@article{b:https://doi.org/10.1029/2021JB023410,
author = {Stefanou, Ioannis and Tzortzopoulos, Georgios},
title = {Preventing Instabilities and Inducing Controlled, Slow-Slip in Frictionally Unstable Systems},
journal = {Journal of Geophysical Research: Solid Earth},
volume = {127},
number = {7},
pages = {e2021JB023410},
doi = {10.1029/2021JB023410},
year = {2022}
}

@ARTICLE{b:Biot-1941,
    Author         = {M. A. Biot},
    Journal        = {Journal of Applied Physics},
    Volume         = {12},
    Number         = {155},
    Pages          = {155-164},
    Title          = {General Theory of Three‐Dimensional Consolidation},
    Year           = {1941} ,
    doi            = {10.1063/1.1712886}
}

@ARTICLE{b:Zastrow-2019,
    Author         = {M. Zastrow},
    Journal        = {Nature},
    Volume         = {},
    Number         = {},
    Pages          = {},
    Title          = {{South Korea accepts geothermal plant probably caused destructive quake}},
    Year           = {2019},
    doi            = {10.1038/d41586-019-00959-4} 
}

@BOOK{b:Orlov-2020,
    Address        = {Cham, Switzerland},
    Author         = {Y. Orlov},
    Publisher      = {Springer International Publishing},
    Title          = {Nonsmooth Lyapunov Analysis in Finite and Infinite Dimensions},    
    Year           = {2020}
}

@ARTICLE{b:Rubinstein-Mahani-2015,
    Author         = {J. L. Rubinstein and A. B. Mahani},
    Journal        = {Seismological Research Letters},
    Volume         = {86},
    Number         = {4},
    Pages          = {1060-1067},
    Title          = {Myths and Facts on Wastewater Injection, Hydraulic Fracturing, Enhanced Oil Recovery, and Induced Seismicity},
    Year           = {2015} ,
    doi            = {10.1785/0220150067}
}

@ARTICLE{b:Stefanou2019,
    Author         = {I. Stefanou},
    Journal        = {Journal of Geophysical Research: Solid Earth},
    Volume         = {124},
    Number         = {8},
    Pages          = {8786-8802},
    Title          = {Controlling anthropogenic and natural seismicity: Insights from active stabilization of the spring-slider
model},
    Year           = {2019},
    doi            = {10.1029/2019JB017847}
}

@BOOK{b:filippov,
    Address        = {Dordrecht, The Netherlands},
    Author         = {A.F. Filippov},
    Publisher      = {Kluwer Academic Publishers},
    Title          = {Differential Equations with Discontinuous Right-hand Sides},
    Year           = {1988}
}

@BOOK{b:Khalil2002,
    Address        = {New Jersey, U.S.A.},
    Author         = {H. Khalil},
    Publisher      = {Prentice Hall},
    Title          = {Nonlinear Systems},
    Year           = {2002}
}

@BOOK{b:Ouhabaz2005,
  author    = {El Maati Ouhabaz},
  title     = {Analysis of Heat Equations on Domains},
  publisher = {Princeton University Press},
  series    = {London Mathematical Society Monographs},
  volume    = {31},
  year      = {2005},
  address   = {USA},
  doi       = {10.1515/9781400826483}
}

@ARTICLE{b:SeeberHornSaturated2019,
  author  = {Richard Seeber and Markus Horn},
  title   = {Guaranteeing Disturbance Rejection and Control Signal Continuity for the Saturated Super-Twisting Algorithm},
  journal = {IEEE Control Systems Letters},
  year    = {2019},
  doi     = {10.1109/LCSYS.2019.2917054}
}

@ARTICLE{b:SeeberReichhartingerConditioned2020,
  author  = {Richard Seeber and Martin Reichhartinger},
  title   = {Conditioned Super-Twisting Algorithm for Systems with Saturated Control Action},
  journal = {Automatica},
  year    = {2020},
  doi     = {10.1016/j.automatica.2020.108921}
}

@misc{KNMI2023,
  author = {{Royal Netherlands Meteorological Institute (KNMI)}},
  title = {Earthquake Catalogue of the Netherlands},
  year = {2023},
  note = {Catalogue used in the data associated with Acosta et al. (2023)},
  url = {https://www.knmi.nl/kennis-en-datacentrum/dataset/aardbevingscatalogus}
}

@article{Wassing2021,
  author  = {Wassing, B. B. T. and Gan, Q. and
             Candela, T. and Fokker, P. A.},
  title   = {Effects of fault transmissivity on the
             potential of fault reactivation and induced
             seismicity: Implications for understanding
             induced seismicity at Pohang EGS},
  journal = {Geothermics},
  volume  = {91},
  pages   = {101976},
  year    = {2021},
  doi     = {10.1016/j.geothermics.2020.101976}
}

@misc{b:Gutierrez-Stefanou-2026_b,
  author = {Guti{\'e}rrez-Oribio, Diego and Stefanou, Ioannis},
  title = {A Control Framework for Induced Seismicity Mitigation in {Groningen} Gas Reservoir},
  year = {2026},
  eprint = {2604.26802},
  archivePrefix = {arXiv},
  doi = {10.48550/arXiv.2604.26802},
  url = {https://arxiv.org/abs/2604.26802}
}

@article{Bourne2014,
  author = {Bourne, S. J. and Oates, S. J. and van Elk, J. and Doornhof, D.},
  title = {A seismological model for earthquakes induced by fluid extraction from a subsurface reservoir},
  journal = {Journal of Geophysical Research: Solid Earth}, volume = {119},
  pages = {8991--9015}, year = {2014}, doi = {10.1002/2014JB011663}
}

@article{WangKrstic2023,
  author = {Wang, Ji and Krstic, Miroslav},
  title = {Event-triggered adaptive control of a parabolic {PDE}--{ODE} cascade with piecewise-constant inputs and identification},
  journal = {IEEE Transactions on Automatic Control}, volume = {68}, number = {9},
  pages = {5493--5508}, year = {2023}, doi = {10.1109/TAC.2022.3227916}
}

@article{Xiao2025,
  author = {Xiao, Yu and Xu, Xiaodong and Luo, Biao and Yang, Chunhua and Gui, Weihua},
  title = {Adaptive event-triggered tracking control for strict-feedback nonlinear {ODE} cascaded coupled hyperbolic {PDE} systems},
  journal = {ISA Transactions}, volume = {164}, pages = {24--33},
  year = {2025}, doi = {10.1016/j.isatra.2025.05.019}
}

@article{JiangWang2025,
  author = {Jiang, Yu-Chen and Wang, Jun-Min},
  title = {Output feedback stabilization of an {ODE}-heat cascade system by neural operator approximations},
  journal = {Systems \& Control Letters}, volume = {204}, pages = {106173},
  year = {2025}, doi = {10.1016/j.sysconle.2025.106173}
}

@article{OrlovEtAl2026,
  author = {Orlov, Yury and Pilloni, Alessandro and Pisano, Alessandro and Usai, Elio},
  title = {Sliding-Mode Boundary Control of an Uncertain Parabolic {PDE}--{ODE} Cascade With Dynamic Actuation and Unmatched Disturbance},
  journal = {IEEE Transactions on Automatic Control}, volume = {71}, number = {6},
  pages = {4073--4080}, year = {2026}, doi = {10.1109/TAC.2025.3648248}
}

@article{GuoEtAl2026,
  author = {Guo, Runsheng and Ma, Min and Qiu, Jianbin and Ji, Wenqiang and Sun, Kangkang},
  title = {Output-feedback adaptive boundary control of scalar hyperbolic {ODE}--{PDE}--{ODE} cascade systems},
  journal = {International Journal of Adaptive Control and Signal Processing},
  volume = {40}, number = {9}, pages = {1807--1817}, year = {2026}, doi = {10.1002/acs.70101}
}

@article{JiangWang2026,
  author = {Jiang, Yu-Chen and Wang, Jun-Min},
  title = {Output regulation of an {ODE}-parabolic {PIDE} cascade system by neural operators},
  journal = {International Journal of Robust and Nonlinear Control},
  year = {2026}, doi = {10.1002/rnc.70714},
  note = {Published online 27 August 2026}
}

\end{document}